\documentclass[pra,two column]{revtex4}
\usepackage{amssymb}
\usepackage{amsmath}

\newtheorem{theorem}{Theorem}
\newtheorem{acknowledgement}[theorem]{Acknowledgement}

\input{tcilatex}
\begin{document}

\title{Photon wave mechanics and position eigenvectors}
\author{Margaret Hawton}
\email{margaret.hawton@lakeheadu.ca}

\begin{abstract}
One and two photon wave functions are derived by projecting the quantum
state vector onto simultaneous eigenvectors of the number operator and a
recently constructed photon position operator [Phys. Rev A \textbf{59}, 954
(1999)] that couples spin and orbital angular momentum. While only the
Landau-Peierls wave function defines a positive definite photon density, a
similarity transformation to a biorthogonal field-potential pair of positive
frequency solutions of Maxwell's equations preserves eigenvalues and
expectation values. We show that this real space description of photons is
compatible with all of the usual rules of quantum mechanics and provides a
framework for understanding the relationships amongst different forms of the
photon wave function in the literature. It also gives a quantum picture of
the optical angular momentum of beams that applies to both one photon and
coherent states. According to the rules of qunatum mechanics, this wave
function gives the probability to count a photon at any position in space.
\end{abstract}

\maketitle
\affiliation{Department of Physics, Lakehead University, Thunder Bay, ON, Canada, P7B 5E1}

\section{ Introduction}

The current interest in entanglement and its application to quantum
information has rekindled the controversy surrounding the photon wave
function \cite{Raymer,LapaireSipe,Rubin,ThePhoton,ScullyBook,FederovEberly}.
It is still unclear what form a real space photon wave function should take,
or if one exists. In the standard formulation of quantum mechanics, the
coordinate space wave function is the projection of the state vector onto an
orthonormal basis of eigenvectors of a Hermitian position operator. It has
been claimed since the early days of quantum mechanics that there is no
position operator for the photon that allows the introduction of a wave
function in this way. Contrary to these claims, we have recently constructed
a photon position operator whose transverse eigenvectors form a real space
basis. Here we will use this basis to obtain a photon wave function that is
compatible with the usual rules of quantum mechanics. We will show that this
clarifies a number of previously unresolved issues regarding the real space
description of one photon and multiphoton states.

In 1933 Pauli \cite{Pauli}\ stated that the nonexistence of a density for
the photon corresponds to the fact that the position of a photon cannot be
associated with an operator in the usual sense. Based on definitions of
center of mass, Pryce found the $\mathbf{k}$-space photon position operator $%
\widehat{\mathbf{r}}_{P}=i\nabla -i\widehat{\mathbf{k}}/2k+\widehat{\mathbf{k%
}}\mathbf{\times S}/k$ where $S_{j}$ are the $3\times 3$\ spin $1$ matrices, 
$\widehat{\mathbf{k}}$ is a unit wave vector, and $\nabla _{j}=\partial
/\partial k_{j}$ \cite{Pryce}. This operator does not have commuting
components which suggests that three spatial coordinates cannot
simultaneously have a definite value. In 1949 Newton and Wigner sought
rotationally invariant localized states and the corresponding position
operators. They were successful for massive particles and zero mass
particles with spin $0$ and $1/2$, but found for photons \textquotedblright
no localized states in the above sense exist\textquotedblright\ \cite%
{NewtonWigner}. This result is widely quoted as a proof of the nonexistence
of a photon position operator. It has been proved that there is no photon
position operator with commuting components that transforms as a vector \cite%
{Jordan80}.

Our position operator \cite{HawtonPO} has commuting components but is not
rotationally invariant and does not transform as a vector \cite%
{HawtonBaylisPO}, and thus it is consistent with the previous work.
Description of a localized state requires a sum over all $\mathbf{k,}$ and a
localized photon can have definite spin in the $\mathbf{k}$-direction, that
is it can have definite helicity, but it cannot have definite spin along any
fixed axis. It is the total angular momentum (AM) that can have a definite
value along some specified direction in space \cite{HawtonBaylisAM}. The
position eigenvectors are not spherically symmetric, instead they have a
vortex structure as is observed for twisted light \cite{AM}. Compared to the
Newton Wigner position operators for which transformation of a particle's
spin and position are separable, the photon position operator must
incorporate an additional unitary transformation that reorients this vortex.

Valid position eigenvectors cannot violate the Hegerfeldt \cite{Hegerfeldt}
and Paley-Wiener \cite{PaleyWiener} theorems based on Fourier transform
theory. Hegerfeldt proved that a positive frequency wave function can be
exactly localized at only one instant in time and interpreted this to imply
a violation of causality. Bialynicki-Birula \cite{BB3} noted that the
Paley-Wiener theorem limits $g\left( x\right) =\int_{0}^{\infty }dkh\left(
k\right) \exp \left( -ikx\right) $ of the form $\exp \left( -Ax^{\gamma
}\right) $ to $\gamma <1.$ He then applied this to separate outgoing and
incoming exponentially localized spherical pulses in three dimensions.
However, their sum is not subject to the exponential localization limit, as
can be seen from the form of the $k$-integral. Position eigenvectors require
a sum over all wave vectors, and thus must be a sum of outgoing and incoming
waves that interfere to give exact localization at a single instant in time,
consistent with the Hegerfeldt theorem.

Maxwell's equations (MEs) are analogous to the Dirac equation when written
in terms of the Riemann-Silberstein (RS) field vector, proportional to $%
\mathbf{E\pm }ic\mathbf{B}$ where $c$ is the speed of light in vacuum, $%
\mathbf{E}$ is the electric field, and $\mathbf{B}$ is the magnetic
induction. This suggests that the photon is an elementary particle like any
other, and that MEs provide a first quantized description of the photon. Use
of the positive frequency RS vector as a photon wave function in vacuum and
in a medium has been thoroughly studied \cite{BB,BB2,Sipe,Raymer}. If a
field $\mathbf{\Psi }^{(1/2)}$ with quantum electrodynamic weighting, $%
k^{1/2},$ is used as wave function, a metric factor $k^{-1}$ is required in
the scalar product. The real space squared norm then goes as$\ \int
d^{3}r\int d^{3}r^{\prime }\mathbf{\Psi }^{(1/2)\ast }\left( \mathbf{r}%
\right) \cdot \mathbf{\Psi }^{(1/2)}\left( \mathbf{r}^{\prime }\right)
/\left\vert \mathbf{r}-\mathbf{r}^{\prime }\right\vert ^{2}$ and thus its
integrand cannot be interpreted as a local number density \cite%
{BB,PikeSarkar}. Since the photon has no mass, it has been suggested that
there is no photon number density, only energy density \cite{Sipe}. Photon
number density based on the Landau-Peierls (LP) wave function $\mathbf{\Psi }%
^{(0)}$ (without the factor $k^{1/2}$) was investigated as early as 1930 
\cite{LandauPeierls,AB}. Its absolute value squared is positive definite but
it has the disadvantage that its relationship to electric current density
and the electromagnetic fields is nonlocal in real space \cite%
{LandauPeierls,AB,BB,Cook}.

Returning to field-like $\mathbf{\Psi }^{(1/2)}$ functions, we will show
here that it is possible to define a biorthonormal basis that gives a local
density by combining the eigenvectors of an operator with those of its
adjoint. This formalism has recently been applied to pseudo-Hermitian
Hamiltonians that possess real spectra \cite{Mostafazadeh}. Such a basis
provides an interesting alternative to explicit inclusion of a metric
operator when working with electromagnetic fields. The density $\mathbf{\Psi 
}^{(1/2)\ast }\left( \mathbf{r}\right) \cdot \mathbf{\Psi }^{(-1/2)}\left( 
\mathbf{r}\right) $ is local, but it not positive definite since it is not
an absolute value squared. Only the LP wave function defines a positive
definite photon density, equal to $\left\vert \mathbf{\Psi }^{(0)}\left( 
\mathbf{r}\right) \right\vert ^{2}$. However, we will show that the
biorthogonal field-potential pair gives the same results in most
calculations.

In the present paper one and two photon wave functions and photon density
will be obtained by projection onto a basis of position eigenvectors. In
Section II the photon position operator will be reviewed and the scalar
product and Hermiticity will be discussed. In Section III the orthonormal
and biorthonormal eigenkets of the position operator will be obtained in the
Heisenberg picture (HP). We will then derive photon wave functions from
quantum electrodynamics (QED) in Section IV by projecting the state vector
onto simultaneous eigenvectors of the photon position, helicity, and number
operators.\ We will discuss MEs, photon wave mechanics, and angular momentum
and beams in Sections V, VI and VII respectively and then conclude.

\section{Position operator}

We start with a discussion of the photon position operator. The procedure
used in \cite{HawtonPO} was to construct an operator with transverse
eigenvectors of definite helicity, $\sigma =\pm 1$. In $\mathbf{k}$-space,
it is reasonable to expect that the transverse function 
\begin{equation}
\psi _{\mathbf{r},\sigma ,j}^{(\alpha )}\left( \mathbf{k}\right) =\left(
\omega _{k}\right) ^{\alpha }e_{\mathbf{k},\sigma ,j}^{(\chi )}\exp \left( -i%
\mathbf{k\cdot r}\right) /\sqrt{V}  \label{PositionEigenvectors}
\end{equation}%
describes a photon located at position $\mathbf{r,}$ where $\omega _{k}=kc$
in vacuum and the parameter $\chi $ will be discussed later in this section.
Subscripts denote eigenvalues and Cartesian components of the vectors $%
\mathbf{\psi }$ and $\mathbf{e}$. Cartesian components are used where it is
necessary to avoid confusing vector notation. The parameter $\alpha $ allows
for both LP and field based wave functions. The position eigenvectors are
electric and/or magnetic fields if $\alpha =1/2,$ the vector potential if $%
\alpha =-1/2,$ or LP wave functions if $\alpha =0.$ This is consistent with
the QED based interpretation that a mode with frequency $\omega _{k}$ has
energy $\hbar \omega _{k}$ so that the square of the fields gives energy
density while the wave function gives number density. The spherical polar
definite helicity unit vectors are 
\begin{equation}
\mathbf{e}_{\mathbf{k},\sigma }^{(0)}=\left( \widehat{\mathbf{\theta }}%
+i\sigma \widehat{\mathbf{\phi }}\right) /\sqrt{2}  \label{e0}
\end{equation}%
\ where $\widehat{\mathbf{\theta }}$ and $\widehat{\mathbf{\phi }}$\ are
unit vectors in the increasing $\theta $ and $\phi $ directions. Periodic
boundary conditions in a finite volume are used here to simplify the
notation, and the limit as $V\rightarrow \infty $ can be taken to calculate
derivatives and perform sums. If the wave function (\ref%
{PositionEigenvectors}) is a position eigenvector it should satisfy the
eigenvector equation%
\begin{equation}
\widehat{\mathbf{r}}^{(\alpha )}\psi _{\mathbf{r},\sigma ,j}^{(\alpha
)}\left( \mathbf{k}\right) =\mathbf{r}\psi _{\mathbf{r},\sigma ,j}^{(\alpha
)}\left( \mathbf{k}\right)  \label{EvectEq}
\end{equation}%
where $\widehat{\mathbf{r}}^{(\alpha )}$ is the $\mathbf{k}$-space
representation of the position operator and its eigenvalues, $\mathbf{r,}$
can be interpreted as photon position.

The operator arrived at in \cite{HawtonPO} using the condition (\ref{EvectEq}%
) is 
\begin{equation}
\widehat{\mathbf{r}}^{(\alpha )}=\widehat{\mathbf{r}}_{P}^{(\alpha )}+S_{%
\mathbf{k}}\widehat{\mathbf{\phi }}\cot \theta /k  \label{rop}
\end{equation}%
where 
\begin{equation}
\widehat{\mathbf{r}}_{P}^{(\alpha )}=iI\nabla -iI\alpha \widehat{\mathbf{k}}%
\mathbf{/}k+\widehat{\mathbf{k}}\mathbf{\times S}/k,  \label{rP}
\end{equation}%
is a generalization of the Pryce operator, $I$ is a $3\times 3$ unit matrix, 
$\left( S_{i}\right) _{jk}=-i\epsilon _{ijk},$ and the component of spin
parallel to $\mathbf{k},$ $S_{\mathbf{k}}=\widehat{\mathbf{\mathbf{k}}}%
\mathbf{\cdot S,}$ extracts the helicity $\sigma .$ The operator $\widehat{%
\mathbf{r}}^{(\alpha )}$ is essentially the usual $\mathbf{k}$-space
position operator, $i\nabla ,$ with terms added to compensate for
differentiation of the unit vectors and $k^{\alpha }$ by $\nabla .$ The term 
$\widehat{\mathbf{k}}\mathbf{\times S}/k$ gives a transverse vector, while $%
S_{\mathbf{k}}\widehat{\mathbf{\phi }}\cot \theta /k$ rotates $\widehat{%
\mathbf{\theta }}$ and $\widehat{\mathbf{\phi }}$ back to their original
orientations, and $-iI\alpha \widehat{\mathbf{k}}\mathbf{/}k$ corrects for
differentiation of $k^{\alpha }.$ It was proved in \cite{HawtonPO} that $%
\widehat{\mathbf{r}}^{(\alpha )}$ has commuting components and satisfies the
other expected commutation relations.

The photon's position coordinates must be real, and this normally implies
that the position operator must be Hermitian. In the LP case the $\mathbf{k}$%
-space inner-product is 
\begin{equation*}
\left\langle \mathbf{\Psi }^{(0)}|\widetilde{\mathbf{\Psi }}%
^{(0)}\right\rangle =\sum_{\mathbf{k},\ j}\Psi {}_{j}^{(0)\ast }\left( 
\mathbf{k}\right) \widetilde{\Psi }_{j}^{(0)}\left( \mathbf{k}\right)
\end{equation*}%
where $\mathbf{\Psi }^{(\alpha )}\left( \mathbf{k}\right) $ and $\widetilde{%
\mathbf{\Psi }}^{(\alpha )}\left( \mathbf{k}\right) $ are any two state
vectors. It can be proved by converting the sum to an integral and
integrating by parts that $\left\langle \mathbf{\Psi }^{(0)}|\widehat{%
\mathbf{r}}^{(0)}\widetilde{\mathbf{\Psi }}^{(0)}\right\rangle =\left\langle 
\widehat{\mathbf{r}}^{(0)}\mathbf{\Psi }^{(0)}|\widetilde{\mathbf{\Psi }}%
^{(0)}\right\rangle $ which implies that $\widehat{\mathbf{r}}^{(0)}$ is
Hermitian. The case $\alpha =1/2$ \ with inner-product 
\begin{equation}
\left\langle \mathbf{\Psi }^{(1/2)}|\widetilde{\mathbf{\Psi }}%
^{(1/2)}\right\rangle =\sum_{\mathbf{k},\ j}k^{-1}\Psi _{j}^{(1/2)\ast
}\left( \mathbf{k}\right) \widetilde{\Psi }_{j}^{(1/2)}\left( \mathbf{k}%
\right)  \label{FieldTheoryIP}
\end{equation}%
was considered in \cite{PikeSarkar} and \cite{HawtonPO}. Integration by
parts in this case requires differentiation of $k^{-1},$which again gives $%
\left\langle \mathbf{\Psi }^{(1/2)}|\widehat{\mathbf{r}}^{(1/2)}\widetilde{%
\mathbf{\Psi }}^{(1/2)}\right\rangle =\left\langle \widehat{\mathbf{r}}%
^{(1/2)}\mathbf{\Psi }^{(1/2)}|\widetilde{\mathbf{\Psi }}^{(1/2)}\right%
\rangle ,$ proving that $\widehat{\mathbf{r}}^{(1/2)}$ is Hermitian based on
the inner-product (\ref{FieldTheoryIP}). This leads to the nonlocal real
space density discussed in the Introduction. Alternatively the inner-product
can be written as%
\begin{equation*}
\left\langle \mathbf{\Psi }^{(1/2)}|\widetilde{\mathbf{\Psi }}%
^{(-1/2)}\right\rangle =\sum_{\mathbf{k},\ j}\Psi _{j}^{(1/2)\ast }\left( 
\mathbf{k}\right) \widetilde{\Psi }_{j}^{(-1/2)}\left( \mathbf{k}\right)
\end{equation*}%
by defining $\widetilde{\mathbf{\Psi }}^{(-1/2)}=\widetilde{\mathbf{\Psi }}%
^{(1/2)}/k,$ thus avoiding explicit inclusion of the factor $k^{-1}$ and the
consequent nonlocal real space density. The expectation value of the
position operator then satisfies $\left\langle \mathbf{\Psi }^{(1/2)}|%
\widehat{\mathbf{r}}^{(-1/2)}\widetilde{\mathbf{\Psi }}^{(-1/2)}\right%
\rangle =\left\langle \widehat{\mathbf{r}}^{(1/2)}\mathbf{\Psi }^{(1/2)}|%
\widetilde{\mathbf{\Psi }}^{(-1/2)}\right\rangle .$ If we apply this to the
localized state $\mathbf{\Psi }^{(\alpha )}=\widetilde{\mathbf{\Psi }}%
^{(\alpha )}=\mathbf{\psi }_{\mathbf{r}^{\prime },\sigma }^{(\alpha )}$ this
proves that the eigenvalue $\mathbf{r}^{\prime }$ is still real. However,
the position operators $\widehat{\mathbf{r}}^{(1/2)}$ and $\widehat{\mathbf{r%
}}^{(-1/2)}=\widehat{\mathbf{r}}^{(1/2)\dagger }$ are not self-adjoint.
Operators such are these are referred to a pseudo-Hermitian by Mostafazadeh 
\cite{Mostafazadeh}. Use of pseudo-Hermitian operators and a biorthonormal
basis is discussed in more detail in the next section.

In \cite{HawtonBaylisPO} the position operator was generalized to allow for
rotation about $\mathbf{k}$ through the Euler angle $\chi \left( \theta
,\phi \right) \ $to give the most general transverse basis, 
\begin{equation}
\mathbf{e}_{\mathbf{k},\sigma }^{\left( \chi \right) }=e^{-i\sigma \chi }%
\mathbf{e}_{\mathbf{k},\sigma }^{(0)}.  \label{e_chi}
\end{equation}%
It was found that the position operator can be written as 
\begin{equation}
\widehat{\mathbf{r}}^{(\alpha )}=D\left( k^{\alpha }i\nabla k^{-\alpha
}\right) D^{-1}  \label{rD}
\end{equation}%
where $D=\exp \left( -iS_{\mathbf{k}}\chi \right) \exp \left( -iS_{3}\phi
\right) \exp \left( -iS_{2}\theta \right) $. Starting from a wave vector
parallel to $\widehat{\mathbf{z}}$ and transverse unit vectors $\widehat{%
\mathbf{x}}$ and $\widehat{\mathbf{y}},$ $D$ rotates $\mathbf{k}$ from the $%
z $-axis to an orientation described by the angles $\theta $ and $\phi ,$
while at the same time rotating the transverse vectors first to $\widehat{%
\mathbf{\theta }}$ and $\widehat{\mathbf{\phi }}$ and then about $\mathbf{k}$
through $\chi .$ For example, when $\widehat{\mathbf{r}}^{(1/2)}$ acts on a
transverse field parallel to $\widehat{\mathbf{\phi }}$ it rotates it to $%
\widehat{\mathbf{y}}$ and divides it by $\sqrt{\omega _{k}},$ then operates
on it with the usual $\mathbf{k}$-space position operator $i\nabla .$ It
then reverses the process by multiplying it by $\sqrt{\omega _{k}}$ and
rotates it back to its original transverse orientation. This allows $%
\widehat{\mathbf{r}}^{(1/2)}$ to extract the position of the photon from the
phase of the coefficient of the transverse unit vector.

The quantum numbers $\left\{ \mathbf{r},\sigma \right\} $ index the basis
states for a given $\chi \left( \theta ,\phi \right) $. The $z$-axis can be
selected for convenience and the choice $\chi =-m\phi $ gives \cite%
{HawtonBaylisAM}%
\begin{eqnarray}
\mathbf{e}_{\mathbf{k},\sigma }^{\left( -m\phi \right) } &=&\frac{\widehat{%
\mathbf{x}}-i\widehat{\mathbf{y}}}{2\sqrt{2}}\left( \cos \theta -\sigma
\right) e^{i\left( m\sigma +1\right) \phi }-\frac{\widehat{\mathbf{z}}}{%
\sqrt{2}}\sin \theta e^{im\sigma \phi }  \notag \\
&&+\frac{\widehat{\mathbf{x}}+i\widehat{\mathbf{y}}}{2\sqrt{2}}\left( \cos
\theta +\sigma \right) e^{i\left( m\sigma -1\right) \phi }.  \label{em}
\end{eqnarray}%
For example, $\chi =-\phi $ ($m=1$) rotates $\widehat{\mathbf{\theta }}$ and 
$\widehat{\mathbf{\phi }}$ back to the $x$ and $y$ axes to give unit vectors
that approach $\left( \widehat{\mathbf{x}}+i\sigma \widehat{\mathbf{y}}%
\right) /\sqrt{2}$ in the $\theta \rightarrow 0$ limit. This is useful in
the description of paraxial beams since the unit vectors describe spin
alone. Their coefficients then describe all of the orbital angular momentum
so that a factor $\exp \left( il_{z}\phi \right) $ implies a $z$-component
of orbital angular momentum equal to $\hbar l_{z}.$

The spin and orbital AM of a photon are not separable beyond the paraxial
approximation. For unit vectors of the form (\ref{em}) the $z$-component of
total angular momentum and photon position operators satisfy \cite%
{HawtonBaylisAM}%
\begin{equation}
\left[ \widehat{J}_{z},\widehat{r}_{k}\right] =i\hbar \epsilon _{zkl}%
\widehat{r}_{l}.  \label{JrCommutation}
\end{equation}%
This is just the usual commutation relation satisfied by a vector operator
and an angular momentum component. Here it implies that photon position
transforms as a vector under rotations about the axis of symmetry of the
localized states. A photon on the $z$-axis satisfies the uncertainty
relation $\Delta J_{z}\Delta r_{k}\geqslant 0.$ Unit vectors of the form (%
\ref{em}) contribute a definite $z$-component of the total AM, consistent
with $\left\{ s_{z},l_{z}\right\} $ equal to$\ \left\{ -1,m\sigma +1\right\} 
$, $\left\{ 0,m\sigma \right\} $ or $\left\{ 1,m\sigma -1\right\} $ with $%
j_{z}=m\sigma ,$ that is total AM has a definite value, but spin and orbital
AM do not.

\section{Position eigenvectors}

Here we will obtain the eigenvectors of the position operators discussed in
Section II. The LP form of the position operator, $\widehat{\mathbf{r}}%
^{(0)},$ is self adjoint, has real eigenvalues, and defines an orthonormal
basis as is usual in quantum mechanics. To obtain QED-like fields as
eigenvectors, the choice $\alpha =1/2$ is required. In this section we will
use the mathematical properties of pseudo-Hermitian operators to obtained a
completeness relation for the field-like photon wave function and
investigate how it is related to the LP wave function. The operators will be
obtained in the HP picture, so time dependence as determined by the
Hamiltonian must also be considered.

We will start with an examination of the expectation values to motivate the
use of the biorthonormal formalism. Any Hermitian operator $\widehat{o}$
satisfies the eigenvector equation $\widehat{o}\left\vert f_{n}\right\rangle
=o_{n}\left\vert f_{n}\right\rangle $ and its eigenvalues, $o_{n}$, are
real. To transform from LP position eigenvectors to fields, multiplication
by $\sqrt{\omega _{k}}$ is required. Assume that $\eta $ is an operator with
positive square root $\rho =\sqrt{\eta }$ which will equal $\sqrt{\omega _{k}%
}$ in the present application. We can write%
\begin{equation*}
\left\langle f_{n}\left\vert \widehat{o}\right\vert f_{m}\right\rangle
=\left\langle f_{n}\left\vert \rho \left( \rho ^{-1}\widehat{o}\rho \right)
\rho ^{-1}\right\vert f_{m}\right\rangle =\left\langle \phi _{n}\left\vert 
\widehat{O}\right\vert \psi _{m}\right\rangle
\end{equation*}%
where $\widehat{O}=\rho ^{-1}\widehat{o}\rho $ is a similarity
transformation, $\left\vert \psi _{m}\right\rangle =\rho ^{-1}\left\vert
f_{m}\right\rangle $ and $\left\vert \phi _{n}\right\rangle =\left(
\left\langle f_{n}\right\vert \rho \right) ^{\dagger }=\rho ^{\dagger }\rho
\left\vert \psi _{n}\right\rangle $. The eigenvector equation becomes $%
\widehat{O}\left\vert \psi _{n}\right\rangle =o_{n}\left\vert \psi
_{n}\right\rangle $ and the eigenvalues and inner-products are preserved.\
If $\rho $ is a unitary operator, that is if $\rho ^{-1}=\rho ^{\dagger },$
then $\widehat{O}^{\dagger }=\widehat{O}$ is Hermitian and $\left\vert \phi
_{n}\right\rangle =\left\vert \psi _{n}\right\rangle .\ $The $\left\vert
\psi _{n}\right\rangle $ and $\left\vert \phi _{n}\right\rangle $
eigenvectors are the same, and the usual quantum mechanical formalism is
obtained. On the other hand, if $\rho $ is a Hermitian operator satisfying $%
\rho ^{\dagger }=\rho $ then $\left\vert \phi _{n}\right\rangle \neq
\left\vert \psi _{n}\right\rangle $ and there are two distinct sets of
eigenvectors. We can deal with this is one of two ways: (1) The metric
operator $\eta =\rho ^{2}$\ can be introduced to give the new inner-product $%
\left\langle \phi _{n}|\eta ^{-1}\phi _{m}\right\rangle $ and work only with
the $\left\vert \phi _{n}\right\rangle $ basis. (2) We can use the
eigenvectors of $\widehat{O}$ \emph{and} the eigenvectors of $\widehat{O}%
^{\dagger }=\eta \widehat{O}\eta ^{-1}$ which are $\left\vert \psi
_{n}\right\rangle $ and $\left\vert \phi _{n}\right\rangle $ respectively.
Since $\left\langle f_{n}|f_{m}\right\rangle =\left\langle f_{n}\left\vert
\rho \rho ^{-1}\right\vert f_{m}\right\rangle $ transforms to $\left\langle
\phi _{n}|\psi _{m}\right\rangle ,$ the eigenvectors $\left\vert \psi
_{m}\right\rangle $ and $\left\vert \phi _{n}\right\rangle $ are
biorthonormal \cite{Fonda}. If there is degeneracy, a biorthonormal basis
can be obtained by defining a complete set of commuting operators (CSCO).

The properties of pseudo-Hermitian operators and biorthonormal bases have
recently been investigated by Mostafazadeh and can be summarized as \cite%
{Mostafazadeh} 
\begin{eqnarray*}
\widehat{O}\left\vert \psi _{n}\right\rangle &=&O_{n}\left\vert \psi
_{n}\right\rangle ,\text{ }\widehat{O}^{\dagger }\left\vert \phi
_{n}\right\rangle =O_{n}^{\ast }\left\vert \phi _{n}\right\rangle , \\
\widehat{O}^{\dagger } &=&\eta \widehat{O}\eta ^{-1},\;\left\langle \phi
_{n}|\psi _{m}\right\rangle =\delta _{n,m},
\end{eqnarray*}%
with the completeness relation 
\begin{equation*}
\sum_{n}\left\vert \psi _{n}\right\rangle \left\langle \phi _{n}\right\vert
=\;\sum_{n}\left\vert \phi _{n}\right\rangle \left\langle \psi
_{n}\right\vert =\widehat{1},
\end{equation*}%
where $\eta $ is a metric operator and $\widehat{1}$\ is the unit operator.
If $\rho =\sqrt{\eta }$ \ exists$,$ 
\begin{equation}
\widehat{o}=\rho \widehat{O}\rho ^{-1}=\rho ^{-1}\widehat{O}^{\dagger }\rho
\label{HermitianO}
\end{equation}%
is self-adjoint and the eigenvectors $O_{n}$ are real. Expectation values
are preserved by the similarity transformation, $\eta $.

To apply this formalism to the photon we take $\eta =\omega _{k}$ and work
in $\mathbf{k}$-space$.$ Then $\widehat{o}=\widehat{\mathbf{r}}^{(0)}$ is
self-adjoint and the operators $\widehat{O}^{\dagger }=\widehat{\mathbf{r}}%
^{(1/2)}$ $\ $and $\widehat{O}=\widehat{\mathbf{r}}^{(-1/2)}$ have the
biorthonormal eigenvectors $\mathbf{\psi }_{\mathbf{r},\sigma
}^{(1/2)}\left( \mathbf{k}\right) $ and $\mathbf{\psi }_{\mathbf{r},\sigma
}^{(-1/2)}\left( \mathbf{k}\right) $ given by Eq. (\ref{PositionEigenvectors}%
) that \ go as $\sqrt{\omega _{k}}$ and $1/\sqrt{\omega _{k}}$ respectively
as required by QED for the electromagnetic fields and the vector potential.
The position operators and their eigenvectors satisfy 
\begin{eqnarray}
\widehat{\mathbf{r}}^{(-\alpha )\dagger } &=&\widehat{\mathbf{r}}^{(\alpha
)},  \label{rAdjoint} \\
\mathbf{\psi }_{\mathbf{r},\sigma }^{(-1/2)}\left( \mathbf{k}\right)
&=&\omega _{k}^{-1/2}\mathbf{\psi }_{\mathbf{r},\sigma }^{(0)}\left( \mathbf{%
k}\right) ,  \label{PsiMinusHalf} \\
\mathbf{\psi }_{\mathbf{r},\sigma }^{(1/2)}\left( \mathbf{k}\right)
&=&\omega _{k}^{1/2}\mathbf{\psi }_{\mathbf{r},\sigma }^{(0)}\left( \mathbf{k%
}\right) ,  \label{PsiHalf}
\end{eqnarray}%
the biorthonormality condition%
\begin{equation}
\sum_{j}\left\langle \psi _{\mathbf{r}^{\prime },\sigma ^{\prime
},j}^{(-\alpha )}|\psi _{\mathbf{r},\sigma ,j}^{(\alpha )}\right\rangle
=\delta ^{3}\left( \mathbf{r}-\mathbf{r}^{\prime }\right) \delta _{\sigma
,\sigma ^{\prime }},  \label{Orthonormal}
\end{equation}%
and the completeness relation 
\begin{equation}
\sum_{\sigma ,j}\int d^{3}r\left\vert \psi _{\mathbf{r},\sigma ,j}^{(\alpha
)}\right\rangle \left\langle \psi _{\mathbf{r},\sigma ,j}^{(-\alpha
)}\right\vert =\widehat{1}.  \label{Complete}
\end{equation}%
Here $\delta ^{3}$ is the $3$-dimensional Dirac $\delta $-function and we
can interchange $\alpha $ and $-\alpha $. The field and the LP operators, $%
\widehat{o}$, are related as%
\begin{equation}
\widehat{O}^{\dagger }=\omega _{k}^{1/2}\widehat{o}\omega _{k}^{-1/2}.
\label{SimilarityTransf}
\end{equation}%
consistent with (\ref{rD}). This transforms the LP position operator $%
\widehat{\mathbf{r}}^{(0)}$ to $\widehat{\mathbf{r}}^{(1/2)},$ introducing
an addition term $-iI\widehat{\mathbf{k}}/2k$. The momentum and angular
momentum \ operators $\hbar \mathbf{k}$ and $\hbar \left( -\mathbf{k}\times
i\nabla +\mathbf{S}\right) $ are unaffected by the similarity transformation
(\ref{SimilarityTransf}). In the angular momentum case this is because $%
\widehat{\mathbf{k}}\times \mathbf{k}=0.$

Time dependence is determined by the Hamiltonian $\widehat{H}+\widehat{H}%
_{0} $ with 
\begin{equation}
\widehat{H}=\sum_{\mathbf{k},\sigma }\hbar \omega _{k}a_{\mathbf{k},\sigma
}^{\dagger }a_{\mathbf{k},\sigma }  \label{Hamiltonian}
\end{equation}%
where the zero point terms $\widehat{H}_{0}=\sum_{\mathbf{k},\sigma }\hbar
\omega _{k}/2$ which are unaffected by the photon state will be omitted
here. The operator $a_{\mathbf{k},\sigma }$ annihilates a photon with wave
vector $\mathbf{k}$ and helicity $\sigma $ and satisfies the commutation
relations $\left[ a_{\mathbf{k},\sigma },a_{\mathbf{k}^{\prime },\sigma
^{\prime }}^{\dagger }\right] =\delta _{\sigma ,\sigma ^{\prime }}\delta _{%
\mathbf{k,k}^{\prime }}$. The operators and their eigenkets are time
dependent in the HP \cite{Sakurai}. Using the unitary time evolution
operator 
\begin{equation}
U\left( t\right) =\exp (-i\widehat{H}t),  \label{TimeEvolution}
\end{equation}%
the HP position operator becomes 
\begin{equation}
\widehat{\mathbf{r}}_{HP}^{(\alpha )}=U^{\dagger }\left( t\right) \widehat{%
\mathbf{r}}^{(\alpha )}U\left( t\right) =\widehat{\mathbf{r}}^{(\alpha
)}+\nabla \omega _{k}t  \label{rHP}
\end{equation}%
with eigenvectors $U^{\dagger }\left( t\right) \left\vert \mathbf{\psi }_{%
\mathbf{r},\sigma }^{(\alpha )}\right\rangle $ with $\left\vert \mathbf{\psi 
}_{\mathbf{r},\sigma }^{(\alpha )}\right\rangle $\ given by Eq. (\ref%
{PositionEigenvectors}) in $\mathbf{k}$-space. The coefficient of $t$ in the
last term of \ (\ref{rHP}) is the photon group velocity.

We can define $1$-photon HP annihilation and creation operators for a photon
with helicity $\sigma $ at position $\mathbf{r}$ and time $t$ as 
\begin{eqnarray}
\widehat{\psi }_{\mathbf{r},\sigma ,j}^{(\alpha )}\left( t\right) &\equiv
&\sum_{\mathbf{k}}\left( \omega _{k}\right) ^{\alpha }e_{\mathbf{k},\sigma
,j}^{(\chi )}a_{\mathbf{k},\sigma }\frac{e^{i\mathbf{k\cdot r}-i\omega _{k}t}%
}{\sqrt{V}},  \label{Annihilation} \\
\widehat{\psi }_{\mathbf{r},\sigma ,j}^{(\alpha )\dagger }\left( t\right)
&\equiv &\sum_{\mathbf{k}}\left( \omega _{k}\right) ^{\alpha }e_{\mathbf{k}%
,\sigma ,j}^{(\chi )\ast }a_{\mathbf{k},\sigma }^{\dagger }\frac{e^{-i%
\mathbf{k\cdot r}+i\omega _{k}t}}{\sqrt{V}}.  \label{Creation}
\end{eqnarray}%
For $\alpha =1/2,$ Eq. (\ref{Annihilation}) implies that the biorthonormal
pairs are related through 
\begin{equation}
\widehat{\mathbf{\psi }}_{\mathbf{r},\sigma }^{(1/2)}\left( t\right) =i\frac{%
\partial \widehat{\mathbf{\psi }}_{\mathbf{r},\sigma }^{(-1/2)}\left(
t\right) }{\partial t}  \label{FieldPotentialEq}
\end{equation}%
analogous to the relationship between the vector potential and the electric
field in the Coulomb gauge. The $1$-photon position eigenkets normalized
according to (\ref{Orthonormal}) are%
\begin{equation}
\left\vert \mathbf{\psi }_{\mathbf{r},\sigma }^{(\alpha )}\left( t\right)
\right\rangle =\widehat{\mathbf{\psi }}_{\mathbf{r},\sigma }^{(\alpha
)\dagger }\left( t\right) \left\vert 0\right\rangle ,  \label{1PhotonEvect}
\end{equation}%
where $\left\vert 0\right\rangle $ is the electromagnetic vacuum state. The
projection of (\ref{1PhotonEvect}) onto the momentum-helicity basis, $%
\left\{ \left\vert \mathbf{k},\sigma \right\rangle \right\} ,$ gives back
Eq. (\ref{PositionEigenvectors}) in the Schr\"{o}dinger picture. The free
space operators for a photon with helicity $\sigma $ satisfy the $\mathbf{r}$%
-space dynamical equation%
\begin{equation}
i\frac{\partial \widehat{\mathbf{\psi }}_{\mathbf{r},\sigma }^{(\alpha
)}\left( t\right) }{\partial t}=\sigma c\nabla \times \widehat{\mathbf{\psi }%
}_{\mathbf{r},\sigma }^{(\alpha )}\left( t\right) .  \label{HelicityOpEq}
\end{equation}%
The annihilation and creation operators satisfy the equal time commutation
relations 
\begin{equation}
\sum_{j}\left[ \widehat{\psi }_{\mathbf{r},\sigma ,j}^{(-\alpha )}\left(
t\right) ,\widehat{\psi }_{\mathbf{r}^{\prime },\sigma ^{\prime
},j}^{(\alpha )\dagger }\left( t\right) \right] =\delta _{\sigma ,\sigma
^{\prime }}\delta ^{3}\left( \mathbf{r}-\mathbf{r}^{\prime }\right) .
\label{Commutation}
\end{equation}%
The Hermitian operator describing the density of photons with helicity $%
\sigma ,$ obtained by averaging over the $\alpha $ and $-\alpha $ forms, is%
\begin{equation}
\widehat{n}_{\sigma }^{(\alpha )}\left( \mathbf{r},t\right) =\frac{1}{2}%
\widehat{\mathbf{\psi }}_{\mathbf{r},\sigma }^{(\alpha )\dagger }\left(
t\right) \cdot \widehat{\mathbf{\psi }}_{\mathbf{r},\sigma }^{(-\alpha
)}\left( t\right) +H.c.\text{.}  \label{DensityOp}
\end{equation}%
The total number operator is%
\begin{equation}
\widehat{N}=\int d^{3}r\widehat{n}^{(\alpha )}\left( \mathbf{r},t\right)
=\sum_{\mathbf{k},\sigma }a_{\mathbf{k},\sigma }^{\dagger }a_{\mathbf{k}%
,\sigma }.  \label{NumberOp}
\end{equation}

An alternative linear polarization basis can be obtained if we define
operators that annihilate a photon state with polarization in the $\widehat{%
\mathbf{\theta }}$ and $\widehat{\mathbf{\phi }}$\ directions as 
\begin{eqnarray}
\widehat{\mathbf{\psi }}_{\mathbf{r}}^{(\alpha )}\left( t\right) &=&\left[ 
\widehat{\mathbf{\psi }}_{\mathbf{r},1}^{(\alpha )}\left( t\right) +\widehat{%
\mathbf{\psi }}_{\mathbf{r},-1}^{(\alpha )}\left( t\right) \right] /\sqrt{2},
\label{PsiPhi} \\
\widehat{\mathbf{\phi }}_{\mathbf{r}}^{(\alpha )}\left( t\right) &=&-i\left[ 
\widehat{\mathbf{\psi }}_{\mathbf{r},1}^{(\alpha )}\left( t\right) -\widehat{%
\mathbf{\psi }}_{\mathbf{r},-1}^{(\alpha )}\left( t\right) \right] /\sqrt{2},
\notag
\end{eqnarray}%
respectively. While the direction of these eigenvectors depends on $\mathbf{k%
}$, they do not rotate in space and time, and in that sense they are
linearly polarized. The inverse transformation is 
\begin{equation}
\widehat{\mathbf{\psi }}_{\mathbf{r},\sigma }^{(\alpha )}\left( t\right) =%
\left[ \widehat{\mathbf{\psi }}_{\mathbf{r}}^{(\alpha )}\left( t\right)
+i\sigma \widehat{\mathbf{\phi }}_{\mathbf{r}}^{(\alpha )}\left( t\right) %
\right] /\sqrt{2}.  \label{toHelicityBasis}
\end{equation}%
In free space 
\begin{eqnarray}
\frac{\partial \widehat{\mathbf{\psi }}_{\mathbf{r}}^{(\alpha )}\left(
t\right) }{\partial t} &=&c\nabla \times \widehat{\mathbf{\phi }}_{\mathbf{r}%
}^{(\alpha )}\left( t\right) ,  \label{PsiPhiOpEq} \\
\frac{\partial \widehat{\mathbf{\phi }}_{\mathbf{r}}^{(\alpha )}\left(
t\right) }{\partial t} &=&-c\nabla \times \widehat{\mathbf{\psi }}_{\mathbf{r%
}}^{(\alpha )}\left( t\right) ,  \notag
\end{eqnarray}%
If $\alpha =0$ these are the operators introduced by Cook \cite{Cook}, while
if $\alpha =1/2$ their dynamics is ME-like.

The localized definite helicity basis states are eigenvectors of a CSCO, so
it is the helicity basis that will be used here. Linearly polarized fields
can be found by taking the sum and difference as in (\ref{PsiPhi}).

\section{Wave function}

In this section we will obtain one and two photon wave functions and photon
density by projection onto the basis of position eigenvectors found in
Section III. This density is a $2$-point correlation function that is based
on the LP or biorthonormal basis, rather than electric fields alone as in
Glauber photodetection theory \cite{Glauber}.

The QED state vector describing a pure state in which the number of photons
and their wave vectors are uncertain can be expanded as 
\begin{eqnarray}
\left\vert \Psi \right\rangle &=&c_{0}\left\vert 0\right\rangle +\sum_{%
\mathbf{k},\sigma }c_{\mathbf{k},\sigma }a_{\mathbf{k},\sigma }^{\dagger
}\left\vert 0\right\rangle  \label{StateVector} \\
&&+\frac{1}{2!}\sum_{\mathbf{k},\sigma ;\mathbf{k}^{\prime },\sigma ^{\prime
}}\sqrt{\mathcal{N}_{\mathbf{k},\sigma ;\mathbf{k}^{\prime },\sigma ^{\prime
}}}c_{\mathbf{k},\sigma ;\mathbf{k}^{\prime },\sigma ^{\prime }}a_{\mathbf{k}%
,\sigma }^{\dagger }a_{\mathbf{k}^{\prime },\sigma ^{\prime }}^{\dagger
}\left\vert 0\right\rangle +..  \notag
\end{eqnarray}%
where $c_{0}=\left\langle 0|\Psi \right\rangle ,$ $c_{\mathbf{k},\sigma
}\equiv \left\langle 0\left\vert a_{\mathbf{k},\sigma }\right\vert \Psi
\right\rangle ,$ $c_{\mathbf{k},\sigma ;\mathbf{k}^{\prime },\sigma ^{\prime
}}\equiv c_{\mathbf{k}^{\prime },\sigma ^{\prime };\mathbf{k},\sigma
}=\left\langle 0\left\vert a_{\mathbf{k},\sigma }a_{\mathbf{k}^{\prime
},\sigma ^{\prime }}\right\vert \Psi \right\rangle ,$ and $\mathcal{N}_{%
\mathbf{k},\sigma ;\mathbf{k}^{\prime },\sigma ^{\prime }}=1+\delta _{%
\mathbf{k},\mathbf{k}^{\prime }}\delta _{\sigma ,\sigma ^{\prime }}$.
Division by $2!$ corrects for identical states obtained when the $\left\{ 
\mathbf{k},\sigma \right\} $ subscripts are permuted while $\sqrt{\mathcal{N}%
}/2$ normalizes doubly occupied states. A more general state requires a
formulation in terms of density matrices that will not be attempted here.

The $1$-photon real space wave function in the helicity basis, equal to the
projection of this state vector onto eigenvectors of $\widehat{\mathbf{r}}%
_{HP}^{(\alpha )}$ as $\left\langle \psi _{\mathbf{r},\sigma ,j}^{(\alpha
)}|\Psi \right\rangle ,$ is 
\begin{equation}
\mathbf{\Psi }_{\sigma }^{(\alpha )}\left( \mathbf{r},t\right) =\sum_{%
\mathbf{k}}c_{\mathbf{k},\sigma }\mathbf{e}_{\mathbf{k,}\sigma }^{(\chi
)}\left( \omega _{k}\right) ^{\alpha }\frac{e^{i\mathbf{k\cdot r}-i\omega
_{k}t}}{\sqrt{V}}  \label{1PhotonPsi}
\end{equation}%
where we have used Eqs. (\ref{Creation}), (\ref{1PhotonEvect}) and (\ref%
{StateVector}). The expansion coefficients depend on the choice of basis,
for example when $\chi \rightarrow \chi +\Delta \chi $ the coefficients $c_{%
\mathbf{k},\sigma }\rightarrow c_{\mathbf{k},\sigma }\exp (-i\sigma \Delta
\chi )$ analogous to gauge changes of the vector potential describing a
magnetic monopole in real space \cite{HawtonBaylisPO}. In any basis the
inner-product $\left\langle \Psi |\Psi \right\rangle =\sum_{\mathbf{k}%
,\sigma }\left\vert c_{\mathbf{k},\sigma }\right\vert ^{2}\equiv \left\vert
c_{1}\right\vert ^{2}$\emph{\ \ }where $\left\vert c_{1}\right\vert ^{2}$ is
the net probability for $1$-photon in state $\left\vert \Psi \right\rangle $%
. The free space $1$-photon dynamical equations mirror the operator Eqs. (%
\ref{FieldPotentialEq}) and (\ref{HelicityOpEq}). They are%
\begin{eqnarray}
i\frac{\partial \mathbf{\Psi }_{\sigma }^{(-1/2)}\left( \mathbf{r},t\right) 
}{\partial t} &=&\mathbf{\Psi }_{\sigma }^{(1/2)}\left( \mathbf{r,}t\right) ,
\label{FieldPotentialWaveEq} \\
i\frac{\partial \mathbf{\Psi }_{\sigma }^{(\alpha )}\left( \mathbf{r}%
,t\right) }{\partial t} &=&\sigma c\nabla \times \mathbf{\Psi }_{\sigma
}^{(\alpha )}\left( \mathbf{r},t\right) .  \notag
\end{eqnarray}

To obtain the $2$-photon wave function we can project $\left\vert \Psi
\right\rangle $ onto the $2$-photon real space basis 
\begin{equation}
\left\vert \psi _{\mathbf{r},\sigma ,j}\left( t\right) ,\psi _{\mathbf{r}%
^{\prime },\sigma ^{\prime },j^{\prime }}\left( t^{\prime }\right)
\right\rangle =\widehat{\psi }_{\mathbf{r},\sigma ,j}^{(\alpha )\dagger
}\left( t\right) \widehat{\psi }_{\mathbf{r}^{\prime },\sigma ^{\prime
},j^{\prime }}^{(\alpha )\dagger }\left( t^{\prime }\right) \left\vert
0\right\rangle  \notag
\end{equation}%
giving%
\begin{equation}
\Psi _{\sigma ,\sigma ^{\prime };j,j^{\prime }}^{(\alpha )}\left( \mathbf{r},%
\mathbf{r}^{\prime };t,t^{\prime }\right) =\left\langle 0\left\vert \widehat{%
\psi }_{\mathbf{r},\sigma ,j}^{(\alpha )}\left( t\right) \widehat{\psi }_{%
\mathbf{r}^{\prime },\sigma ^{\prime },j^{\prime }}^{(\alpha )}\left(
t^{\prime }\right) \right\vert \Psi \right\rangle .  \label{2Photon}
\end{equation}%
Use of Eq. (\ref{Creation}) and $\left[ a_{\mathbf{k},\sigma },a_{\mathbf{k}%
^{\prime },\sigma ^{\prime }}^{\dagger }\right] =\delta _{\mathbf{k},\mathbf{%
k}^{\prime }}\delta _{\sigma ,\sigma ^{\prime }}$ to evaluate $\left\langle
0\right\vert \widehat{\psi }_{\mathbf{r},\sigma ,j}^{(\alpha )}\left(
t\right) \widehat{\psi }_{\mathbf{r}^{\prime },\sigma ^{\prime },j^{\prime
}}^{(\alpha )}\left( t^{\prime }\right) a_{\mathbf{k},\sigma ^{\prime \prime
}}^{\dagger }a_{\mathbf{k}^{\prime },\sigma ^{\prime \prime \prime
}}^{\dagger }\left\vert 0\right\rangle $ then gives 
\begin{eqnarray}
\Psi _{\sigma ,\sigma ^{\prime };j,j^{\prime }}^{(\alpha )}\left( \mathbf{r},%
\mathbf{r}^{\prime };t,t^{\prime }\right) &=&\frac{1}{2!V}\sum_{\mathbf{k}%
,\sigma ;\mathbf{k}^{\prime },\sigma ^{\prime }}\sqrt{\mathcal{N}_{\mathbf{k}%
,\sigma ;\mathbf{k}^{\prime },\sigma ^{\prime }}}  \label{2PhotonPsi} \\
&&\times c_{\mathbf{k},\sigma ;\mathbf{k}^{\prime },\sigma ^{\prime }}\left(
\omega _{k}\omega _{k^{\prime }}\right) ^{\alpha }  \notag \\
&&\times \left[ e_{\mathbf{k},\sigma ,j}^{(\chi )}e_{\mathbf{k}^{\prime
},\sigma ^{\prime },j^{\prime }}^{(\chi )}e^{i\mathbf{k\cdot r}-i\omega
_{k}t}e^{i\mathbf{k}^{\prime }\mathbf{\cdot r}^{\prime }-i\omega _{k^{\prime
}}t^{\prime }}\right.  \notag \\
&&\left. +e_{\mathbf{k}^{\prime },\sigma ^{\prime },j}^{(\chi )}e_{\mathbf{k}%
,\sigma ,j^{\prime }}^{(\chi )}e^{i\mathbf{k\cdot r}^{\prime }-i\omega
_{k}t^{\prime }}e^{i\mathbf{k}^{\prime }\mathbf{\cdot r}-i\omega _{k^{\prime
}}t}\right]  \notag
\end{eqnarray}%
which becomes a $2$-photon wave function if we set $t^{\prime }=t$. A
separate symmetrization step is not required since its symmetric form is a
direct consequence of the commutation relations satisfied by the photon
annihilation and creation operators.

To obtain an $n$-photon basis the creation operator can be applied to the
vacuum $n$ times with each occurrence having different parameters $\mathbf{r,%
}$ $\sigma ,$ and $j$. The state vector can then be projected onto this
basis to give the $n$-photon term. The result is the symmetric $n$-photon
real space function 
\begin{equation}
\Psi _{\left\{ m\right\} }^{(\alpha )}\left( \mathbf{r},..,\mathbf{r}%
^{[n-1]};t,..,t^{[n-1]}\right) =\dprod\limits_{m=0}^{n-1}\left\langle \psi
_{m}^{(\alpha )}\right\vert \left\vert \Psi \right\rangle  \label{nPhoton}
\end{equation}%
\ where $\left\vert \psi _{m}^{(\alpha )}\right\rangle $ is a short hand for 
$\left\vert \psi _{\mathbf{r}^{[m]},\sigma ^{\lbrack m]},j^{[m]}}^{(\alpha
)}\left( t^{[m]}\right) \right\rangle $ and $m$ represents the $m^{th}$ set
of variables, quantum numbers and components $\left\{ \mathbf{r}%
^{[m]},t^{[m]},\sigma ^{\lbrack m]},j^{[m]}\right\} $. Generally the $n$%
-photon states provide more information than can be measured. Instead the
real space helicity $\sigma $ photon density, equal to the expectation value
of the number density operator, (\ref{DensityOp}), can be defined as 
\begin{eqnarray}
n_{\sigma }^{(\alpha )}\left( \mathbf{r},t\right) &=&\left\langle \Psi
\left\vert \widehat{n}_{\sigma }\left( \mathbf{r},t\right) \right\vert \Psi
\right\rangle  \label{Correlation} \\
&=&\frac{1}{2}\sum_{j}\left\langle \Psi \left\vert \widehat{\psi }_{\mathbf{r%
},\sigma ,j}^{(\alpha )\dagger }\left( t\right) \widehat{\psi }_{\mathbf{r}%
,\sigma ,j}^{(-\alpha )}\left( t\right) \right\vert \Psi \right\rangle +c.c.
\notag
\end{eqnarray}%
The $0$-photon contribution to $n$ is $0$, while the $1$-photon contribution
is 
\begin{equation}
n_{\sigma }^{(\alpha )}\left( \mathbf{r},t\right) =\frac{1}{2}\mathbf{\Psi }%
_{\sigma }^{(\alpha )\ast }\left( \mathbf{r},t\right) \cdot \mathbf{\Psi }%
_{\sigma }^{(-\alpha )}\left( \mathbf{r},t\right) +c.c..
\label{PhotonDensity}
\end{equation}%
For the $2$-photon state (\ref{2Photon}), substitution of (\ref{Commutation}%
) gives%
\begin{eqnarray*}
n_{\sigma }^{(\alpha )}\left( \mathbf{r},t\right) &=&\frac{1}{2}\sum_{\sigma
^{\prime };j,j^{\prime }}\int d^{3}r^{\prime }\Psi _{\sigma ,\sigma ^{\prime
};j,j^{\prime }}^{(\alpha )\ast }\left( \mathbf{r},\mathbf{r}^{\prime
};t,t\right) \\
&&\times \Psi _{\sigma ,\sigma ^{\prime };j,j^{\prime }}^{(-\alpha )}\left( 
\mathbf{r},\mathbf{r}^{\prime };t,t\right) +c.c.,
\end{eqnarray*}%
implying that unobserved photons are summed over. A similar argument can be
applied to each $n$-photon term. Photons are noninteracting particles and
the existence of a photon density is consistent with Feynman's conclusion
the photon probability density can be interpreted as particle density \cite%
{Feynman,Inagaki}.

The bases obtained here provide a real space description of the multiphoton
state that \textquotedblright encodes the maximum total knowledge describing
the system\textquotedblright\ as discussed in Ref. \cite{Rubin}. The
electric field wave function used in \cite{ScullyBook,LapaireSipe} or RS
vectors in \cite{BB,Raymer}\ by themselves do not provide a basis, and this
is the root of the criticism of \cite{LapaireSipe} made in \cite{Rubin}. The 
$2$-photon wave function (\ref{2PhotonPsi}) is symmetric in agreement with 
\cite{LapaireSipe,Raymer}.

\section{Maxwell's equations}

In this section we will show that MEs can be obtained from QED in two
distinct ways. The first is the conventional approach of calculating the
expectation value of operators with all modes populated as coherent states.
The fields obtained in this way are real and they cannot be interpreted as
wave functions. The second approach is to project the state vector onto the
position eigenvectors obtained when a field operator acts on the vacuum
state to give fields proportional to the $1$-photon wave function components
in Section IV.

If the multipolar Hamiltonian is used, the displacement is purely photonic,
while the vector potential will include photon and matter contributions \cite%
{CT}. The vector potential operator is a sum over positive and negative
frequencies, photon and matter parts, and both helicities. We can define%
\begin{eqnarray}
\widehat{\mathbf{A}}\left( \mathbf{r},t\right) &=&\widehat{\mathbf{A}}%
^{(+)}\left( \mathbf{r},t\right) +\widehat{\mathbf{A}}^{(-)}\left( \mathbf{r}%
,t\right) ,\   \label{A} \\
\widehat{\mathbf{A}}^{(+)}\left( \mathbf{r},t\right) &=&\widehat{\mathbf{A}}%
_{p}^{(+)}\left( \mathbf{r},t\right) +\widehat{\mathbf{A}}_{m}^{(+)}\left( 
\mathbf{r},t\right)  \notag \\
\widehat{\mathbf{A}}_{p}^{(+)}\left( \mathbf{r},t\right) &=&\widehat{\mathbf{%
A}}_{1}^{(+)}\left( \mathbf{r},t\right) +\widehat{\mathbf{A}}%
_{-1}^{(+)}\left( \mathbf{r},t\right) ,  \notag
\end{eqnarray}%
where $\widehat{\mathbf{A}}^{(-)}=\widehat{\mathbf{A}}^{(+)\dagger }$ and
the subscripts $m$ and $p$ denote matter and photon parts respectively. The
electric field and magnetic induction are then given by 
\begin{eqnarray}
\widehat{\mathbf{E}} &=&-\partial \widehat{\mathbf{A}}/\partial t-\nabla
\phi ,  \label{EandB} \\
\ \widehat{\mathbf{B}} &=&\nabla \times \widehat{\mathbf{A}}.  \notag
\end{eqnarray}%
In the presence matter of with polarization operator $\widehat{\mathcal{P}}$
\ and magnetization $\widehat{\mathcal{M}}$ the displacement and magnetic
field operators are 
\begin{eqnarray}
\widehat{\mathbf{D}} &=&\epsilon _{0}\widehat{\mathbf{E}}+\widehat{\mathcal{P%
}},  \label{FieldOps} \\
\widehat{\mathbf{H}} &=&\widehat{\mathbf{B}}/\mu _{0}-\widehat{\mathcal{M}},
\notag
\end{eqnarray}%
where\ SI units are used, $\epsilon _{0}$ is the permittivity and $\mu _{0}$
the permeability of vacuum, and $c=1/\sqrt{\epsilon _{0}\mu _{0}}$.

The momentum conjugate to the vector potential is $-\widehat{\mathbf{D}}%
_{\perp }$ where $\widehat{\mathbf{D}}_{\perp }$ is the transverse part of
the displacement operator \cite{CT,Sipe}. These operators satisfy canonical
commutation relations. Since $\widehat{\mathbf{\psi }}_{\mathbf{r},\sigma
}^{(-\alpha )}$ and $\widehat{\mathbf{\psi }}_{\mathbf{r},\sigma }^{(\alpha
)\dagger }$ satisfy (\ref{Commutation}) we can choose 
\begin{eqnarray}
\widehat{\mathbf{A}}_{\sigma }^{(+)}\left( \mathbf{r},t\right) &=&\sqrt{%
\frac{\hbar }{2\epsilon _{0}}}\widehat{\mathbf{\psi }}_{\mathbf{r},\sigma
}^{(-1/2)}\left( t\right) ,  \label{A+} \\
\widehat{\mathbf{D}}_{\perp ,\sigma }^{(+)}\left( \mathbf{r},t\right) &=&i%
\sqrt{\frac{\hbar \epsilon _{0}}{2}}\widehat{\mathbf{\psi }}_{\mathbf{r}%
,\sigma }^{(1/2)}\left( t\right) .  \notag
\end{eqnarray}%
This is equivalent to the usual QED expansion of $\widehat{\mathbf{A}}$ and $%
\widehat{\mathbf{D}}$ and thus is consistent with the operator MEs%
\begin{eqnarray}
\nabla \cdot \widehat{\mathbf{B}} &=&0,\ \nabla \times \widehat{\mathbf{E}}=-%
\frac{\partial \widehat{\mathbf{B}}}{\partial t},  \label{OpMEs} \\
\nabla \cdot \widehat{\mathbf{D}} &=&\rho ,\ \nabla \times \widehat{\mathbf{H%
}}=\mathbf{j+}\frac{\partial \widehat{\mathbf{D}}}{\partial t},  \notag
\end{eqnarray}%
where $\rho $ and $\mathbf{j}$ are the free charge and current densities. In
free space $\widehat{\mathbf{D}}_{\perp ,\sigma }^{(+)}/\sqrt{\epsilon _{0}}%
=i\sigma \widehat{\mathbf{B}}_{\sigma }^{(+)}/\sqrt{\mu _{0}}=\widehat{%
\mathbf{F}}_{\sigma }^{(+)}/\sqrt{2}=i\sqrt{\hbar /2}\widehat{\mathbf{\psi }}%
_{\mathbf{r},\sigma }^{(1/2)}$ where the RS operator is $\widehat{\mathbf{F}}%
_{\sigma }^{(+)}=\widehat{\mathbf{D}}_{\sigma }^{(+)}/\sqrt{2\epsilon _{0}}+%
\widehat{\mathbf{B}}_{\sigma }^{(+)}/\sqrt{2\mu _{0}}$ as defined in \cite%
{BB}.

Coherent states are the most classical, and they can be used to establish a
connection between QED and the real Maxwell fields. Following
Cohen-Tannoudji et. al. \cite{CT} the complex Fourier transforms of the
classical field vectors, 
\begin{equation*}
\mathcal{V}_{\mathbf{k}}\left( t\right) =\int d^{3}r\mathbf{V}\left( \mathbf{%
r},t\right) \frac{\exp \left( -i\mathbf{k\cdot r}\right) }{\sqrt{V}},
\end{equation*}%
and the normal variables,%
\begin{equation*}
\mathbf{\gamma }_{\mathbf{k}}\left( t\right) =-i\sqrt{\frac{\epsilon _{0}}{%
2\hbar \omega _{k}}}\left[ \mathcal{E}_{\mathbf{k}}^{\bot }\left( t\right) -c%
\widehat{\mathbf{k}}\times \mathcal{B}_{\mathbf{k}}\left( t\right) \right] ,
\end{equation*}%
can be defined. For a coherent state with the photon occupancy of mode $%
\left\{ \mathbf{k},\sigma \right\} $ described by the complex parameter $%
\gamma _{\mathbf{k},\sigma },$ the average photon number is $n_{\mathbf{k}%
,\sigma }=\left\vert \gamma _{\mathbf{k},\sigma }\right\vert ^{2}$ and\ the
probability amplitude for $n$-photons is $\exp \left( -\left\vert \gamma _{%
\mathbf{k},\sigma }\right\vert ^{2}/2\right) \gamma _{\mathbf{k},\sigma
}^{n}/\sqrt{n!}.$ The quasi-classical coherent state is a Gaussian wave
packet that oscillates without deformation and with relative number
uncertainty $\Delta n_{\mathbf{k},\sigma }/n_{\mathbf{k},\sigma }=1/$ $%
\left\vert \gamma _{\mathbf{k},\sigma }\right\vert .$ In the limit of
infinite photon number the electric and magnetic fields oscillate in a well
defined way as do the solutions to the classical MEs. Thus%
\begin{eqnarray}
\mathbf{A}_{coh}^{\perp (+)}\left( \mathbf{r},t\right) &=&\left\langle
\left\{ \gamma _{\mathbf{k},\sigma }\right\} \left\vert \widehat{\mathbf{A}}%
_{p}^{\perp (+)}\left( \mathbf{r},t\right) \right\vert \left\{ \gamma _{%
\mathbf{k},\sigma }\right\} \right\rangle  \label{PosCohA} \\
&=&\sum_{\mathbf{k},\sigma }\sqrt{\frac{\hbar }{2\epsilon _{0}\omega _{k}}}%
\mathbf{\gamma }_{\mathbf{k},\sigma }\mathbf{e}_{\mathbf{k,}\sigma }^{(\chi
)}\frac{e^{i\mathbf{k\cdot r}-i\omega _{k}t}}{\sqrt{V}},  \notag \\
\mathbf{A}_{coh}^{\perp }\left( \mathbf{r},t\right) &=&\mathbf{A}%
_{coh}^{\perp (+)}\left( \mathbf{r},t\right) +c.c.  \label{CohA}
\end{eqnarray}%
and the fields derived from it behave classically in the large photon number
limit$.$

It is also possible to derive one photon positive frequency MEs from QED.
Since it is still widely believed that there is no position basis for the
photon, this result is new. We can define the $1$-particle states $%
\left\vert \mathbf{V}_{\mathbf{r},\sigma }\right\rangle =\widehat{\mathbf{V}}%
_{\mathbf{r},\sigma }^{(-)}\left\vert g,0\right\rangle $ with $\widehat{%
\mathbf{V}}^{(-)}=\widehat{\mathbf{V}}^{(+)\dagger }$ and $\mathbf{V}%
^{(+)}=\sum_{\sigma }\mathbf{V}_{\sigma }^{(+)}$ for any field operator $%
\widehat{\mathbf{V}}$ such that 
\begin{equation}
\mathbf{V}_{\sigma }^{(+)}\left( \mathbf{r},t\right) =\left\langle g,0|%
\widehat{\mathbf{V}}_{\mathbf{r},\sigma }^{(+)}|\Psi \right\rangle .
\label{V+}
\end{equation}%
This can be viewed as the projection of the photon-matter state vector state
onto the $n=1$ term of number-position-helicity basis. In the ground state $%
\left\vert 0\right\rangle $ both the EM field and any matter present are in
their lowest energy configurations. The operator $\widehat{\mathbf{V}}^{(-)}$
creates $1$-particle that can be a photon or a material excitation.\ Since
the space and time dependence originates entirely in the field operators,
these functions satisfy ME dynamics. The $1$-photon MEs are, using (\ref%
{OpMEs}),%
\begin{eqnarray}
\nabla \cdot \mathbf{B}^{(+)} &=&0,\ \nabla \times \mathbf{E}^{(+)}=-\frac{%
\partial \mathbf{B}^{(+)}}{\partial t},  \label{MEs} \\
\nabla \cdot \mathbf{D}^{(+)} &=&\rho ^{(+)},\ \nabla \times \mathbf{H}%
^{(+)}=\mathbf{j^{(+)}+}\frac{\partial \mathbf{D}^{(+)}}{\partial t}.  \notag
\end{eqnarray}%
Projection of the state vector onto a basis of $1$-photon position
eigenvectors results in intrinsically positive frequency electric and
magnetic fields defined by (\ref{V+}) that satisfy MEs. They can be
manipulated to give any of the commonly used forms of MEs.

A wave equation can be obtained from (\ref{MEs}) in the usual way to give 
\begin{eqnarray}
\frac{1}{c^{2}}\frac{\mathbf{\partial }^{2}\mathbf{E}^{(+)}}{\partial t^{2}}%
+\nabla \times \nabla \times \mathbf{E}^{(+)} &=&-\mu _{0}\frac{\partial }{%
\partial t}\left( \frac{\partial \mathcal{P}^{(+)}}{\partial t}\right.
\label{WaveWithSource} \\
&&\left. +\mathbf{\nabla \times }\mathcal{M}^{(+)}+\mathbf{j}^{(+)}\right) .
\notag
\end{eqnarray}%
The terms on the right hand side are the polarization, magnetic and external
contributions to the time derivative of the current density\ \cite{CT}. If
there is no magnetization or external current and the polarization is linear
and isotropic we can write $\mathcal{P}=\epsilon _{0}\chi \left( k\right) 
\mathbf{E}$ which can be combined with the $\partial ^{2}\mathbf{E}%
^{(+)}/\partial t^{2}$ term. Writing $\epsilon \left( k\right) =\epsilon _{0}%
\left[ 1+\chi \left( k\right) \right] $ the angular frequency in (\ref%
{Hamiltonian}) is $\omega _{k}=kc\sqrt{1+\chi \left( k\right) }.$ Analogous
to the creation of an excitation of the electromagnetic field (a photon) by $%
\widehat{\mathbf{D}}^{(-)},$ the polarization operator $\widehat{\mathcal{P}}%
^{(-)}$ creates a matter excitation. Energy can be transferred between
matter and the electromagnetic fields, so the matter and EM modes are
coupled. Self-consistent solution of the matter-photon dynamical equations
gives the polariton frequencies $\omega _{k}$ that determine time dependence.

The energy, linear momentum and angular momentum of the free electromagnetic
field are conserved. Their densities and associated currents satisfy
continuity equations of the form $\partial \rho /\partial t+\nabla \cdot 
\mathbf{j}=0.$ This can be verified using MEs, and the steps in this
derivation are still valid if we replace the products of classical real
fields with Hermitian linear combinations of products of operators. For
example, the current describing the flow of energy density $\left\langle 
\widehat{\mathbf{D}}^{(-)}\cdot \widehat{\mathbf{D}}^{(+)}/2\epsilon _{0}+%
\widehat{\mathbf{B}}^{(-)}\cdot \widehat{\mathbf{B}}^{(+)}/2\mu
_{0}\right\rangle $ is $c^{2}$ times the linear momentum density 
\begin{equation}
\mathbf{P}\left( \mathbf{r},t\right) =\frac{1}{2}\left\langle \Psi
\left\vert \widehat{\mathbf{D}}^{(-)}\times \widehat{\mathbf{B}}^{(+)}-%
\widehat{\mathbf{B}}^{(-)}\times \widehat{\mathbf{D}}^{(+)}\right\vert \Psi
\right\rangle .  \label{P}
\end{equation}%
Together with their associated current densities the components of $\mathbf{P%
}$ also satisfy continuity equations which implies that $\int d^{3}r\mathbf{P%
}\left( \mathbf{r},t\right) $ is a constant of the motion. If $\left\vert
\Psi \right\rangle $ is a $1$-photon state $\left\langle \Psi \right\vert 
\widehat{\mathbf{D}}^{(-)}\times \widehat{\mathbf{B}}^{(+)}\left\vert \Psi
\right\rangle =\left\langle \Psi \right\vert \widehat{\mathbf{D}}%
^{(-)}\left\vert 0\right\rangle \times \left\langle 0\right\vert \widehat{%
\mathbf{B}}^{(+)}\left\vert \Psi \right\rangle $ so that%
\begin{equation}
\mathbf{P}\left( \mathbf{r},t\right) =\frac{1}{2}\left[ \mathbf{D}%
^{(-)}\left( \mathbf{r},t\right) \times \mathbf{B}^{(+)}\left( \mathbf{r}%
,t\right) +c.c.\right]  \label{1photonP}
\end{equation}%
with the fields derived using (\ref{EandB}), (\ref{FieldOps}), and (\ref{V+}%
). For a coherent state, the quasi-classical expectation value $\left\langle
\left\{ \gamma _{\mathbf{k},\sigma }\right\} \right\vert \widehat{\mathbf{D}}%
\times \widehat{\mathbf{B}}\left\vert \left\{ \gamma _{\mathbf{k},\sigma
}\right\} \right\rangle \neq \mathbf{D}_{coh}\times \mathbf{B}_{coh}$ for
small $\left\vert \gamma _{\mathbf{k},\sigma }\right\vert .$ However (\ref{P}%
) can be evaluated exactly using $a_{\mathbf{k},\sigma }\left\vert \gamma _{%
\mathbf{k},\sigma }\right\rangle $ to give%
\begin{equation}
\mathbf{P}\left( \mathbf{r},t\right) =\frac{1}{2}\left[ \mathbf{D}%
_{coh}^{(-)}\left( \mathbf{r},t\right) \times \mathbf{B}_{coh}^{(+)}\left( 
\mathbf{r},t\right) +c.c.\right]  \label{CoherentP}
\end{equation}%
with $\mathbf{A}_{coh}^{\perp (+)}$ given by (\ref{PosCohA}). In either case
the angular momentum density is%
\begin{equation}
\mathbf{J}\left( \mathbf{r},t\right) =\mathbf{r}\times \mathbf{P}\left( 
\mathbf{r},t\right) .  \label{Jconserved}
\end{equation}

We are now in a position to compare the classical and quantum fields and
densities. Eq. (\ref{CohA}) describes real fields that are the expectation
values for coherent quantum states. Expectation values do not describe $1$%
-photon states since in this case the expectation values of the field
operators are zero. Instead, it is projection onto a basis of position
eigenvectors that gives $1$-photon positive frequency fields, proportional
to components of the wave function. For $1$-photon and coherent states
momentum density can be written as a cross product of fields as in (\ref%
{1photonP}) and (\ref{CoherentP}). Eq. (\ref{P}) can be used to interpolate
between these two extreme cases.

The density $\mathbf{D}^{(-)}\times \mathbf{B}^{(+)}$ can be rewritten as 
\cite{CT}%
\begin{eqnarray*}
\mathbf{D}^{(-)}\times \mathbf{B}^{(+)} &=&\mathbf{D}^{(-)}\times \left(
\nabla \times \mathbf{A}^{(+)}\right) \\
&=&\sum_{j=1}^{3}D_{j}^{(-)}\nabla A_{j}^{(+)}-\left( \mathbf{D}^{(-)}\cdot
\nabla \right) \mathbf{A}^{(+)}.
\end{eqnarray*}%
Its first term, equal to%
\begin{equation*}
\sum_{j=1}^{3}D_{j}^{(-)}\nabla A_{j}^{(+)}=\frac{1}{2}\sum_{j=1}^{3}\Psi
_{j}^{(1/2)\ast }\left( i\hbar \nabla \right) \Psi _{j}^{(-1/2)},
\end{equation*}%
is the integrand in the expectation value of the real space momentum
operator $-i\hbar \nabla $. The last term, $\left( \mathbf{D}^{(-)}\cdot
\nabla \right) \mathbf{A}^{(+)},$ also contributes to the flow of energy
density and has important consequences. It is responsible for the spin term
in the AM (\ref{Jconserved}). This can be seen by writing%
\begin{eqnarray*}
-\mathbf{r\times }\left( \mathbf{D}^{(-)}\cdot \nabla \right) \mathbf{A}%
^{(+)} &=&\mathbf{D}^{(-)}\times \mathbf{A}^{(+)} \\
&&-\left( \mathbf{D}^{(-)}\cdot \nabla \right) \left( \mathbf{r\times A}%
^{(+)}\right)
\end{eqnarray*}%
where $\mathbf{a\times b}=-i\left( \mathbf{a\cdot S}\right) \mathbf{b}$ gives%
\begin{equation*}
\mathbf{D}^{(-)}\times \mathbf{A}^{(+)}=\frac{1}{2}\sum_{j=1}^{3}\Psi
_{j}^{(1/2)\ast }\hbar \mathbf{S}\Psi _{j}^{(-1/2)}.
\end{equation*}%
Since $\nabla .\mathbf{D}^{(-)}=\rho ^{(-)},$ the last term contributes $%
\int d^{3}r\mathbf{r\times \rho A}^{(+)}$ to $\int d^{3}r\mathbf{J}\left( 
\mathbf{r},t\right) $ after integration by parts which is zero in the
absence of free charge.

\section{ Photon wave mechanics}

In this section we will discuss first quantized photon quantum mechanics.\
For definiteness we will refer to the Barut-Marlin rules for Schr\"{o}dinger
and Dirac particles stated in \cite{Barut} as: (a) A basis for the space of
wave functions, which describe all the possible states of a particle, is
defined by a wave equation. (b) A inner-product is defined in the space of
the wave functions. (c) Expressions for the probability density and
probability current are found. They should form a 4-vector whose divergence
vanishes. The expression for the probability density should be positive
definite. (d) Operators which correspond to measurements are defined, in
particular, momentum and position operators. (e) The eigenfunction of the
operators, normalized to 1 (in the case of discrete spectrum) or a $\delta $%
-function (in the case of a continuous function), are found. (f) The
position operator, defined in (d), and the inner-product, defined in (b),
uniquely determine an expression for the probability density. The theory is
consistent only if this uniquely determined expression is identical with the
one defined in (c) to satisfy a continuity equation. This is a consistency
test.

In brief, these rules apply to the $\mathbf{r}$-space wave mechanics of a
single free photon in free space in the following sense: (a) Solutions to (%
\ref{FieldPotentialWaveEq}), 
\begin{equation}
i\partial \mathbf{\Psi }_{\sigma }^{(\alpha )}\left( \mathbf{r},t\right)
/\partial t=\sigma c\nabla \times \mathbf{\Psi }_{\sigma }^{(\alpha )}\left( 
\mathbf{r},t\right) ,  \label{HelicityEq}
\end{equation}%
include positive and negative frequencies. The negative frequency solution
can be eliminated on physical grounds \cite{Inagaki,BB2}, thus cutting the
Hilbert space in half as is done for solutions to the Dirac equation \cite%
{Barut}. (b) The inner-product of the wave functions describing states $%
\left\vert \widetilde{\Psi }\right\rangle $ and $\left\vert \Psi
\right\rangle ,$%
\begin{equation}
\left\langle \widetilde{\Psi }^{(\alpha )}|\Psi ^{(-\alpha )}\right\rangle
=\sum_{\sigma }\int d^{3}r\widetilde{\mathbf{\Psi }}_{\sigma }^{(\alpha
)\dagger }\left( \mathbf{r},t\right) \cdot \mathbf{\Psi }_{\sigma
}^{(-\alpha )}\left( \mathbf{r},t\right) ,  \label{ScalarProduct}
\end{equation}%
exists and is invariant under similarity transformations between $\alpha
=1/2\ $and $\alpha =0.$ (c) The real number and current densities obtained
by averaging the $\alpha $ and $-\alpha $ densities%
\begin{eqnarray}
n^{(\alpha )}\left( \mathbf{r},t\right) &=&\frac{1}{2}\sum_{\sigma }\mathbf{%
\Psi }_{\sigma }^{(\alpha )\ast }\cdot \mathbf{\Psi }_{\sigma }^{(-\alpha
)}+c.c.,  \label{Density} \\
\mathbf{j}^{(\alpha )}\left( \mathbf{r},t\right) &=&-\frac{i\sigma c}{2}%
\sum_{\sigma }\mathbf{\Psi }_{\sigma }^{(\alpha )\ast }\times \mathbf{\Psi }%
_{\sigma }^{(-\alpha )}+c.c.,  \notag
\end{eqnarray}%
satisfy the continuity equation 
\begin{equation}
\frac{\partial n^{(\alpha )}\left( \mathbf{r},t\right) }{\partial t}+\nabla
\cdot \mathbf{j}^{(\alpha )}\left( \mathbf{r},t\right) =0.
\label{Continuity}
\end{equation}%
This can be verified using the wave equation. The density $%
n^{(0)}=\sum_{\sigma }\left\vert \Psi _{\sigma }^{(0)}\right\vert ^{2}$ is
positive definite, while $\left[ n^{(1/2)},\mathbf{j}^{(1/2)}\right] $ is a $%
4$-vector that can be written as the contraction of second rank EM field
tensors with $4$-potentials. (d) The momentum operator is $\hbar \mathbf{k}$
and the position operator is given by Eq. (\ref{rop}). (e) The eigenvectors
of these operators are $\delta $-function normalized according to (\ref%
{Orthonormal}). (f) The position operator and inner-product give the density 
$\frac{1}{2}\left\langle \mathbf{\psi }_{\mathbf{r},\sigma }^{(\alpha
)}|\Psi \right\rangle ^{\ast }\left\langle \mathbf{\psi }_{\mathbf{r},\sigma
}^{(-\alpha )}|\Psi \right\rangle +c.c.$. Some of these points will now be
discussed in more detail.

\ Both positive and negative frequency solutions of the wave equation are
mathematically allowed. The classical solutions are real, and real waves do
not satisfy a continuity equation or allow a probability interpretation \cite%
{Bohm}. It has been argued by Inagaki for LP wave functions that the
negative frequency solutions with momentum in opposite direction to the wave
propagation should be discarded from the physical photon state \cite{Inagaki}%
. A similar case is made by Bialynicki-Birula for elimination of the
negative frequency fields in field-like wave functions \cite{BB,BB2}.

As with MEs the photon wave equations can be written in a number of
equivalent ways, and this will be considered next to allow comparison with
the existing photon wave function literature.\ The six component wave
function%
\begin{equation}
\Psi _{hel}^{(\alpha )}=\left( 
\begin{array}{c}
\mathbf{\Psi }_{1}^{(\alpha )} \\ 
\mathbf{\Psi }_{-1}^{(\alpha )}%
\end{array}%
\right)  \label{helWaveFtn}
\end{equation}%
in the helicity basis and%
\begin{equation}
\Psi _{lin}^{(\alpha )}=\left( 
\begin{array}{c}
\mathbf{\Psi }^{(\alpha )} \\ 
\mathbf{\Phi }^{(\alpha )}%
\end{array}%
\right)  \label{linWaveFunction}
\end{equation}%
in the linear polarization basis can be defined. The Schr\"{o}dinger
equation is then, using (\ref{HelicityEq}) and (\ref{PsiPhiOpEq}) with $%
\nabla \times \mathbf{a}=-i\left( \mathbf{S\cdot \nabla }\right) \mathbf{a},$
\begin{equation}
i\frac{\partial }{\partial t}\left( 
\begin{array}{c}
\mathbf{\Psi }_{1}^{(\alpha )} \\ 
\mathbf{\Psi }_{-1}^{(\alpha )}%
\end{array}%
\right) =c\left( 
\begin{array}{cc}
-i\mathbf{S\cdot }\nabla & 0 \\ 
0 & i\mathbf{S\cdot }\nabla%
\end{array}%
\right) \left( 
\begin{array}{c}
\mathbf{\Psi }_{1}^{(\alpha )} \\ 
\mathbf{\Psi }_{-1}^{(\alpha )}%
\end{array}%
\right)  \label{BBWaveEq}
\end{equation}%
in the helicity basis and%
\begin{equation}
i\frac{\partial }{\partial t}\left( 
\begin{array}{c}
\mathbf{\Psi }^{(\alpha )} \\ 
\mathbf{\Phi }^{(\alpha )}%
\end{array}%
\right) =c\left( 
\begin{array}{cc}
0 & \mathbf{S\cdot }\nabla \\ 
-\mathbf{S\cdot }\nabla & 0%
\end{array}%
\right) \left( 
\begin{array}{c}
\mathbf{\Psi }^{(\alpha )} \\ 
\mathbf{\Phi }^{(\alpha )}%
\end{array}%
\right)  \label{CookWaveEq}
\end{equation}%
in the linear polarization basis. If $\alpha =1/2,$ (\ref{BBWaveEq}) is of
the form considered by Bialynicki-Birula and Sipe \cite{BB,Sipe}, while if $%
\alpha =0$ (\ref{CookWaveEq}) is the form used by Inagaki \cite{Inagaki}.
However, (\ref{BBWaveEq}) and (\ref{CookWaveEq}) themselves imply that
either the helicity or the linear polarization basis can be used in
combination with field-like $\alpha =1/2$ wave functions \emph{or} LP $%
\alpha =0$ wave functions. \ The operator on the right hand sides of (\ref%
{BBWaveEq}) and (\ref{CookWaveEq}) is the real space $1$-photon Hamiltonian.

The density $i\epsilon _{0}\mathbf{E}\cdot \mathbf{A}/\hbar $ has appeared
before in the classical context and in applications to beams.
Cohen-Tannoudji et. al. \cite{CT} transform the classical electromagnetic
angular momentum as 
\begin{eqnarray}
\mathbf{J} &=&\epsilon _{0}\int d^{3}r\mathbf{r\times }\left( \mathbf{%
E\times B}\right)  \label{CTJ} \\
&=&\epsilon _{0}\int d^{3}r\left[ \sum_{i=1}^{3}E_{i}\left( \mathbf{r\times
\nabla }\right) A_{i}+\mathbf{E\times A}\right]  \notag
\end{eqnarray}%
by requiring that the fields go to zero sufficiently quickly at infinity.
Although this looks like an expectation value, the fields are classical. In
a discussion of optical beams, van Enk and Nienhuis \cite{vanEnk} separate
monochromatic fields into their positive and negative frequency parts using 
\begin{equation*}
\mathbf{V=}\left[ \mathbf{V}^{(+)}\exp \left( -i\omega t\right) +\mathbf{V}%
^{(-)}\exp \left( i\omega t\right) \right] /\sqrt{2}
\end{equation*}%
and obtain for total field linear momentum and AM 
\begin{eqnarray}
\mathbf{P} &=&-i\int d^{3}r\left[ \sum_{i=1}^{3}D_{i}^{(+)\ast }\left( i%
\mathbf{\nabla }\right) A_{i}^{(+)}\right] ,  \label{vanEnkP} \\
\mathbf{J} &=&-i\int d^{3}r\left[ \sum_{i=1}^{3}D_{i}^{(+)\ast }\left( -%
\mathbf{r\times }i\mathbf{\nabla +S}\right) A_{i}^{(+)}\right] .
\label{vanEnkJ}
\end{eqnarray}%
Here we have assumed the absence of matter in writing $\mathbf{D=\epsilon }%
_{0}\mathbf{E}$, substituted $\mathbf{A}^{(+)}=i\omega \mathbf{D}^{(+)}$,
and changed the notation a bit for consistency with the present work. These
are classical expressions, but terms at frequency $2\omega $ do not
contribute to the total momentum and angular momentum, $\mathbf{P}$ and $%
\mathbf{J}$ \cite{SimmonsGuttman}. They look like the expectations values of
the linear and angular momentum operators that would be obtained using the
biorthonormal wave function pair $\sqrt{\epsilon _{0}/\hbar }\mathbf{A}%
_{photon}^{(+)\bot }$ and $-i\mathbf{D}^{(+)}/\sqrt{\epsilon _{0}\hbar }$.
The number operator $i\widehat{\mathbf{D}}^{(-)}\cdot \widehat{\mathbf{A}}%
^{(+)}/2\hbar +H.c.$ was shown previously to be the zeroth component of a
four-vector obtained by contraction of the second rank EM field tensor with
the four-potential $(\phi ,\mathbf{A)}$ \cite{HawtonMelde}. This demonstates
that the biorthonormal basis is of value for comparison with the existing
literature.

It was noted in Section III that the biorthonormal inner-product is
equivalent to the use of a metric operator. Using (\ref{FieldPotentialEq})
and $\widehat{H}=\hbar kc$ in $\mathbf{k}$-space and substituting $\widehat{H%
}$ for $i\partial /\partial t$ the inner-product (\ref{ScalarProduct}) can
be written as%
\begin{eqnarray*}
\left\langle \widetilde{\Psi }|\Psi \right\rangle &=&\sum_{\sigma ,j}\int 
\frac{d^{3}k}{kc}\widetilde{\Psi }_{\sigma ,j}^{(1/2)\ast }\left( \mathbf{k}%
,t\right) \Psi _{\sigma ,j}^{(1/2)}\left( \mathbf{k},t\right) \\
&=&\sum_{\sigma ,j}\int d^{3}r\widetilde{\Psi }_{\sigma ,j}^{(1/2)\ast
}\left( \mathbf{r},t\right) \widehat{H}^{-1}\Psi _{\sigma ,j}^{(1/2)}\left( 
\mathbf{r},t\right)
\end{eqnarray*}%
as in \cite{PikeSarkar,Chakrabarti}.

The number density is the expectation value of the number density operator (%
\ref{DensityOp}) as discussed in Section IV. The $\alpha =\pm 1/2$ wave
function pair gives a real local $1$-photon density $n^{(1/2)},$ but this
density is not positive definite. This can be seen from the following
example: If $\left\vert \Psi \right\rangle $ is a $1$-photon state that
includes only wave vectors $\mathbf{k}_{1}$ and $\mathbf{k}_{2}=\mathbf{k}%
_{1}+\Delta \mathbf{k}$ both with helicity $\sigma $ where $c_{\mathbf{k}%
_{1},\sigma }=c_{\mathbf{k}_{2},\sigma }=1/\sqrt{2}$ then 
\begin{eqnarray*}
n^{(1/2)} &=&\{1+\frac{1}{2}\left( \sqrt{k_{1}/k_{2}}+\sqrt{k_{2}/k_{1}}%
\right) \\
&&\times \cos \left[ \Delta \mathbf{k}\cdot \mathbf{r}-(k_{1}-k_{2})ct)%
\right] \}/V.
\end{eqnarray*}%
The cosine term can exceed the spatially uniform time independent term due
to the $\sqrt{k}$ factors, leading to negative values. If $k_{2}\approx
k_{1} $, $n^{(1/2)}$ is approximately equal to the positive definite
density, $n^{(0)},$ however only the LP wave function satisfies the positive
definite requirement exactly.

It thus appears that LP wave functions are essential to a probability
interpretation. Field-like wave functions can be obtained from the LP wave
function by a similarity transformation, and thus are equivalent to it for
the calculation of expectation values. The operators given by Eq. (\ref%
{PsiPhi}) in the $\alpha =0$ case are identical to the operators examined by
Cook. The equations that they satisfy differ from those for $\mathbf{D}$ and 
$\mathbf{B}$ only in that their relationship to charge and current sources
is nonlocal. The LP number density has been criticized \cite{AB,BB,Sipe},
but its scalar analog, obtained by taking Fourier transforms of the Schmidt
modes, has recently been applied to spontaneous emission of a photon by an
atom and spontaneous parametric down-conversion \cite{Eberly,FederovEberly}.
For narrowband superpositions of plane wave states the distinction between
the LP and field-like form of the wave function has no observable
consequences \cite{FederovEberly}.

The operator (\ref{Annihilation}) creates basis states that lead to the
orthogonal transverse $1$-photon wave function $\Psi _{hel}^{(\alpha
)}\left( \mathbf{r},t\right) =\left[ \mathbf{\Psi }_{1}^{(\alpha )},\mathbf{%
\Psi }_{-1}^{(\alpha )}\right] $ in the helicity basis. The wave function
components $\Psi ^{(-1/2)}$ are proportional to the vector potential, while $%
\Psi ^{(1/2)}$ is related to EM fields. Contraction of \ the second rank
field tensor $F^{\mu \nu }=\partial ^{\nu }A^{\mu }-\partial ^{\mu }A^{\nu }$
with the $4$-potential as $F^{\mu \nu \ast }A_{\nu }$ gives a $4$-vector 
\cite{HawtonMelde}. Thus $\left[ n^{(1/2)},\mathbf{j}^{(1/2)}\right] $ is a $%
4$-vector and photon density is its zeroth component.

In the linear polarization basis the density operators are%
\begin{eqnarray}
\widehat{n}^{(\alpha )}\left( \mathbf{r},t\right) &=&\frac{1}{2}\left[ 
\widehat{\mathbf{\psi }}_{\mathbf{r}}^{(\alpha )\dagger }\cdot \widehat{%
\mathbf{\psi }}_{\mathbf{r}}^{(-\alpha )}+\widehat{\mathbf{\phi }}_{\mathbf{r%
}}^{(\alpha )\dagger }\cdot \widehat{\mathbf{\phi }}_{\mathbf{r}}^{(-\alpha
)}+H.c.\right] ,  \label{linDensityOp} \\
\widehat{\mathbf{j}}^{(\alpha )}\left( \mathbf{r},t\right) &=&\frac{1}{2}%
\left[ \widehat{\mathbf{\psi }}_{\mathbf{r}}^{(\alpha )\dagger }\times 
\widehat{\mathbf{\phi }}_{\mathbf{r}}^{(-\alpha )}-\widehat{\mathbf{\phi }_{%
\mathbf{r}}}^{(-\alpha )\dagger }\times \widehat{\mathbf{\psi }}_{\mathbf{r}%
}^{(\alpha )}+H.c.\right] ,  \notag
\end{eqnarray}%
with $\widehat{\mathbf{\psi }}_{\mathbf{r}}^{(\alpha )}$ and $\widehat{%
\mathbf{\phi }}_{\mathbf{r}}^{(\alpha )}$ given by (\ref{PsiPhi}). Mandel
and Wolf noted the convenience of a photon number operator, equal to $%
\widehat{\mathbf{\psi }}_{\mathbf{r}}^{(0)\dagger }\cdot \widehat{\mathbf{%
\psi }}_{\mathbf{r}}^{(0)}$ in the present notation, to the theory of photon
counting for an arbitrary quantum state \cite{Mandel}. Cook sought detector
independent photon density and current operators that satisfy a continuity
equation. His operators are just (\ref{linDensityOp}) if we take $\alpha =0$%
. Inagaki reformulated Cook's theory in terms of conventional quantum
mechanics \cite{Inagaki}. These authors discuss the restrictions imposed by
photon nonlocalizability, but the existence of a basis of position
eigenvectors makes this unnecessary here. Our operators describe microscopic
densities, and there is no restriction based on wave length.

The Lorentz transformation properties of the $\alpha =0$ photon annihilation
operators in the linear polarization basis were also considered by Cook \cite%
{Cook2}. He concluded that their continuity equation is covariant in the
sense that is is related to the field vectors in the same way in all
reference frames. \ The Hamiltonian, momentum, angular momentum, and Lorentz
transformation operators must conform to the Poincar\'{e} algebra. Since the
position operator generates a change in particle momentum, the boost
operator is closely related to the position operator. For a free photon in $%
\mathbf{k}$-space the Lorentz operator corresponding to the $\alpha =1/2$
case is \cite{Chakrabarti}%
\begin{equation*}
\widehat{\mathbf{K}}^{(1/2)}=k\left( i\nabla \right) +\widehat{\mathbf{k}}%
\times \mathbf{S}.
\end{equation*}%
where $\widehat{\mathbf{K}}^{(-1/2)}=\widehat{\mathbf{K}}^{(1/2)\dagger }$.
Using (\ref{SimilarityTransf}) this gives%
\begin{equation*}
\widehat{\mathbf{K}}^{(0)}=k^{-1/2}\widehat{\mathbf{K}}^{(1/2)}k^{1/2}
\end{equation*}%
\ for the LP boost operator which incorporates the similarity
transformation. In $\mathbf{k}$-space this is simple, but in $\mathbf{r}$%
-space it is non-local as discussed by Cook \cite{Cook2}.

It is stated, in \cite{Fring} for example, that "the non-Hermitian
formulation is in most cases a mere change of metric of a well posed
Hermitian problem. Nonetheless, .. , it has been successfully argued that
the non-Hermitian formalism is often more natural and simplifies
calculations." These comments apply here. The choice of $\alpha $ does not
affect expectation values, the inner-product, and the existence of a wave
equation and a continuity equation. Only the number and current densities
themselves are affected. The field and LP bases can be viewed as alternative
descriptions of the photon state. For most purposes fields are more closely
related to the physics, but the LP basis is needed if the band width is
large and photon number density is required.

According to the general rules of quantum mechanics, for a $1$--photon state
the probability that a photon with helicity $\sigma $ will be found at
position $\mathbf{r}$ at time $t$ is $\left\vert \Psi _{\sigma }^{(0)}\left( 
\mathbf{r},t\right) \right\vert ^{2}\mathbf{.}$ More generally the photon
number density is the expectation value of the number density operator, $%
n_{\sigma }^{(0)}\left( \mathbf{r},t\right) \approx n_{\sigma
}^{(1/2)}\left( \mathbf{r},t\right) $ given by (\ref{Correlation}). Glauber 
\cite{Glauber} defined an ideal photodector to be of negligible size with a
frequency-independent photoabsorption probability. An ideal photon counting
detector also has a quantum efficiency of $\eta =1$, that is any photon
reaching the detector is counted. A detector with all of these
characteristics measures photon position. Consider a $1$-photon pulse
travelling in the $z$-direction that is normally incident on a detector of
thickness $\Delta z$ and area $\Delta A.$ The probability that a photon is
present in this detector, and hence that it is counted, is $n_{\sigma
}^{(\alpha )}\left( \mathbf{r},t\right) \Delta A\Delta z$. In Glauber theory
the count rate is $dn_{G}/dt\propto \left\langle \Psi \left\vert \widehat{%
\mathbf{E}}^{(-)}\left( \mathbf{r},t\right) \cdot \widehat{\mathbf{E}}%
^{(+)}\left( \mathbf{r},t\right) \right\vert \Psi \right\rangle $ where $%
\left( dn_{G}/dt\right) \Delta z/c$ is the probability the photon is counted
during the time that it takes to traverse the detector. Since $n_{\sigma
}^{(1/2)}=i\epsilon _{0}\mathbf{E}_{\sigma }^{(-)}\cdot \mathbf{A}_{\sigma
}^{(+)}/\hbar +c.c$ where $A_{\sigma }^{(+)}\approx -iE_{\sigma }^{(+)}/%
\overline{\omega }$ for most beams available in the laboratory, the
predictions of the present photon number based theory and Glauber
photodetection theory are usually indistinguishable.

The number based theory has the advantage that the probability is
normalizable, for example the probability to count one photon in a $1$%
-photon state in the whole of space using an array of detectors with $\eta
=1 $. The Glauber form of the count rate is based on the transition
probability, however there are advocates for a photon number density
approach, even within conventional photon counting theory. Mandel noted that
"there are many problems in quantum optics, particularly those concerned
with photoelectric measurements of the field, which are most conveniently
treated with the help of an operator representing the number of photons" 
\cite{Mandel}. Mandel and Wolf based their general photon counting theory on
a photon number operator \cite{MandelWolf}. Cook observed that there is no
universal proportionality constant that relates photon flux to $p^{G}$, and
thus the prevailing theory of photoelectron counting fails to provide a
complete description of photon transport \cite{Cook} . He proposed a
modified photodetection theory based on photon number. A photon density $%
n_{\sigma }^{(0)}\left( \mathbf{r},t\right) ,$ equal to the probability
density to count a photon at $\mathbf{r}$ at time $t,$ is consistent with
Cook's arguments \emph{and} with the rules of quantum mechanics.

\section{Angular momentum and beams}

The physical interpretation of the position eigenvectors in \cite%
{HawtonBaylisAM} involving AM was motivated by the recent experimental and
theoretical work on optical vortices. These vortices are spiral phase ramps
described by fields that go as $\exp \left( il_{z}\varphi \right) $ and in
experiments appear as annular rings around a dark center. It can be seen by
inspection of (\ref{em}) that the localized states must have orbital AM, and
this implies a vortex structure that is affected by the choice of $\chi $.
Taking helicity $\sigma =1$ to give a concrete example, we can first take $%
m=0$ in (\ref{em}) to give the spherical polar vectors $\left( \widehat{%
\mathbf{\theta }}+i\sigma \widehat{\mathbf{\phi }}\right) /\sqrt{2}$ with
total AM $0$.\ At $\theta =0$ there is spin AM $\hbar $ and the orbital AM
is $-\hbar $, while at $\theta =\pi $ the spin and orbital AM are $-\hbar $
and $\hbar $ respectively. If instead we choose $m=1$, the $\theta =0$
orbital AM is $0,$ but at $\theta =\pi $ it is $2\hbar $. For a localized
state the vortex has not been eliminated, it has just been moved. Thus an
understanding of optical AM is essential to the physical picture of the
localized basis states that are used here to obtain the photon wave function.

Theoretically, the simplest beams with orbital AM are the nondiffracting
Bessel beams (BBs), and these beams are closely related to our localized
states. They satisfy MEs and have definite frequency, $ck_{0},$ and a
definite wave vector, $k_{z},$ along the propagation direction$.$ It then
follows that the $\mathbf{k}$-space transverse wave vector magnitude $%
k_{\bot }=\sqrt{k_{0}^{2}-k_{z}^{2}},$ and the angle $\theta =\tan
^{-1}\left( k_{\bot }/k_{z}\right) $ also have definite values for BBs.
Cylindrical symmetry is achieved by weighting all $\phi $ equally with a
phase factor $\exp \left( im\phi \right) .$ When Fourier transformed to $%
\mathbf{r}$-space the modes go as $\exp \left( -ik_{0}ct+il_{z}\varphi
+ik_{z}z\right) J_{l_{z}}\left( k_{\perp }r\right) $ where $l_{z}=m$ and $%
m\pm 1$ in (\ref{em}), $J_{l_{z}}$ are Bessel functions, $\varphi $ the real
space azimuthal angle, and $r$ is the perpendicular distance from the beam
axis \cite{Hacyan}. If we select $\chi =0$ so that the $\mathbf{k}$-space
unit vectors are $\widehat{\mathbf{\phi }}$ and $\widehat{\mathbf{\theta }}$
in the linear polarization basis, $\mathbf{B}$ is transverse to $\widehat{%
\mathbf{z}}$ for the $\widehat{\mathbf{\theta }}$ mode and $\mathbf{E}$ is
transverse for the $\widehat{\mathbf{\phi }}$ mode and the linearly
polarized modes can be called transverse magnetic (TM) and transverse
electric (TE) respectively.

The Bessel functions have a sinusoidal dependence on $k_{\bot }r,$ and this
implies that the BBs are standing waves that are a sum of incoming and
outgoing waves. If integrated over $k_{\bot }$ the resulting wave is
localized on the $z$-axis at some instant in time that can be defined as $%
t=0 $. Localization of beams in this way is discussed in \cite{Saari,Bagan}.
If the BBs are then integrated over $k_{z},$ the result is equivalent to a
sum over all wave vectors, and states localized in three dimensions are
obtained. But note that this $k_{z}$ sum includes waves travelling in the
positive and negative $\widehat{\mathbf{z}}$-directions. According to the
Paley-Weiner theorem, $\int_{0}^{\infty }dk_{z}$ does not allow exact
localization, but this restriction does not apply to an integral over all
positive and negative values. Position is not a constant of the motion, and
localized states can exist only for an instant in time. Exactly localized
states in free space are not physically possible because they require
infinite energy. However, our primary concern here is with the use of
localized basis states for calculation of the photon wave function, and we
do not require that these basis states have a physical realization.

The real space mathematical description of beams used to interpret the AM
experiments is usually based on the classical energy, linear momentum, and
angular momentum densities. Here, with a basis of position eigenvectors in
hand that leads to a wave function for a photon in an arbitrary state, we
are in a position to consider the real space description of the AM of beams
from a quantum mechanical perspective. The $\alpha =1/2$ wave function is a
solution to MEs, and any derivation based on MEs can be adapted to the $1$%
-photon case. The expansion of vector potential in \cite{Bagan} that leads
to paraxial fields to a first approximation can be applied to allow
application of our formalism to the paraxial beams that are used in most
optical experiments. Localized states do not exist within the paraxial
approximation, and the paraxial approximation cannot be applied to the
position eigenvectors.

A paraxial beam propagating in the $\widehat{\mathbf{z}}$-direction with
frequency $\omega ,$ helicity $\sigma ,$ and $z$-component of orbital AM $%
\hbar l_{z}$ can be described in cylindrical polar coordinates by the vector
potential \cite{Allen92}%
\begin{equation}
\mathbf{A}^{(+)}\left( \mathbf{r},t\right) =\frac{1}{2}\left( \widehat{%
\mathbf{x}}+i\sigma \widehat{\mathbf{y}}\right) u\left( r\right) \exp \left[
il_{z}\varphi +ik_{z}\left( z-ct\right) \right] .  \label{Abeam}
\end{equation}%
This vector potential is equivalent to the wave function $\mathbf{\Psi }%
_{\sigma ^{\prime }}\left( \mathbf{r},t\right) =\delta _{\sigma ,\sigma
^{\prime }}\sqrt{2\epsilon _{0}/\hbar }\mathbf{A}^{(+)}\left( \mathbf{r}%
,t\right) $. The $z$-component of the time average of the classical AM
density, equal to $\frac{1}{2}\mathbf{r\times }\left( \mathbf{D}^{\ast }%
\mathbf{\times B+D\times B}^{\ast }\right) ,$ is then found to be%
\begin{equation}
J_{z}\left( r\right) =\epsilon _{0}\left[ \omega l_{z}\left\vert u\left(
r\right) \right\vert ^{2}-\frac{1}{2}\omega \sigma r\frac{\partial
\left\vert u^{2}\left( r\right) \right\vert }{\partial r}\right] .
\label{Jbeam}
\end{equation}%
It equals the $z$-component of the AM density (\ref{Jconserved}) with
momentum density given (\ref{1photonP}) or (\ref{CoherentP}) without the
need for time averaging. Thus (\ref{Jbeam}) can be interpreted as a quantum
mechanical AM density that is valid for coherent and $1$-photon states,
while (\ref{P}) interpolates between these two cases.

The first term of (\ref{Jbeam}) is consistent with orbital AM $\hbar l_{z}$
per photon since the photon density given by (\ref{PhotonDensity}) reduces
to $n^{(1/2)}\left( \mathbf{r},t\right) =\epsilon _{0}\omega \left\vert
u\left( r\right) \right\vert ^{2}/\hbar .$ The last term of (\ref{Jbeam})
does not look like photon spin density. The most paradoxical case is a plane
wave, as discussed in \cite{AllenPadgett}. For example a wave function
proportional to $\left( \widehat{\mathbf{x}}+i\sigma \widehat{\mathbf{y}}%
\right) \exp \left( ikz-i\omega t\right) $ implies linear momentum $\hbar k%
\widehat{\mathbf{z}}$ per photon\ and hence no $z$-component of AM. But we
know that such a beam describes a stream of photons each with spin AM $\hbar
\sigma .$ It was observed in 1936 by Beth \cite{Beth} that a circularly
polarized beam can cause a disk to rotate, so the beam really does carry AM
that it can transfer to the disk. The AM of this beam resides in its edges,
as can be seen from Eq. (\ref{Jbeam}). A new edge is created if the disk
intercepts part of the beam and this reduces the AM of the beam, allowing
the conservation of total AM \cite{SimmonsGuttman}. This is analogous to the
continuum description of a dielectric where it is know that the medium is
composed of atoms, but a continuum description of a uniformly polarized
dielectric results only is a surface charge. An even closer analogy exists
between spin AM and a continuous magnetic medium where a current in
individual molecules reduces to a macroscopic current at the edges of the
medium.

In quantum mechanics operators describe observables and their eigenvalues
are the possible results of a measurement. While spin and orbital AM are in
general not separable, the choice $\chi =-\phi $ in Eq. (\ref{e_chi}) gives
unit vectors $\left( \widehat{\mathbf{x}}+i\sigma \widehat{\mathbf{y}}%
\right) /\sqrt{2}$ in the paraxial limit which implies spin quantum number $%
s_{z}=\pm 1.$ The wave function (\ref{Abeam}) is an eigenvector of $\widehat{%
S}_{z}$ with eigenvalue $s_{z}\ $and of $\widehat{L}_{z}=-i\hbar \partial
/\partial \varphi $ with eigenvalue $\hbar l_{z}$ where $\varphi $ the real
space azimuthal angle. The latter orbital AM is equivalent to linear
momentum $\hbar l_{z}/r.$ For this definite helicity state only one term in
the photon density (\ref{Density}) contributes. The probability to detect
this photon is $n^{(0)}\left( \mathbf{r},t\right) \cong n^{(1/2)}\left( 
\mathbf{r},t\right) ,$ where these field-potential and LP densities are
essentially equal for a paraxial beam. For a coherent state the expansion
coefficients $c_{\mathbf{k},\sigma }$ in the $1$-photon wave function (\ref%
{1PhotonPsi}) are replaced with the amplitudes $\gamma _{\mathbf{k},\sigma }$%
.\ Small absorbing particles placed in these beams are essentially
photodetectors that conserve AM by spinning about their centers of mass and
rotating around the beam axis while they absorb photons \cite{AM2}. The
photon number density gives the probability to absorb a photon which carries
with it spin AM $\hbar s_{z}$ and orbital AM $\hbar l_{z}$. For transparent
particles the situation is more complicated, since re-emission should also
be considered.

\section{Conclusion}

We have derived one and two photon wave functions from QED by projecting the
state vector onto the eigenvectors of a photon position operator. Largely
because it is still widely believed that there is no position operator, this
is the first time that a photon wave function has been obtained in this way.
The two photon wave function is symmetric, in agreement with \cite{Raymer}
and \cite{LapaireSipe}. While only the LP wave function gives a positive
definite photon density, field-like wave functions are widely used and are
more convenient in many applications. Also, they given energy momentum and
angular momentum density as in (\ref{1photonP}) for example. In the
field-like helicity basis the wave function pair is%
\begin{eqnarray}
\mathbf{\Psi }_{\sigma }^{(-1/2)}\left( \mathbf{r},t\right) &=&\sqrt{\frac{%
2\epsilon _{0}}{\hbar }}\mathbf{A}_{\sigma }^{(+)}\left( \mathbf{r},t\right)
,  \label{RSField} \\
\mathbf{\Psi }_{\sigma }^{(1/2)}\left( \mathbf{r},t\right) &=&-i\sqrt{\frac{2%
}{\hbar \epsilon _{0}}}\mathbf{D}_{\sigma }^{(+)}\left( \mathbf{r},t\right) .
\notag
\end{eqnarray}%
The wave function components $\mathbf{\Psi }_{\sigma }^{(\alpha )}$ are
given by Eq. (\ref{1PhotonPsi}). For definite helicity fields in free space, 
$\mathbf{B}^{(+)}$ and the Reimann-Silberstein field vector are just
proportional to $\mathbf{D}^{(+)},$ and thus are equivalent to it. The
linear polarization basis of TM and TE fields can be obtained by taking the
sum and difference of the definite helicity modes as in (\ref{PsiPhi}). The
photon density is (\ref{Density}) 
\begin{equation}
n_{\sigma }^{(\alpha )}\left( \mathbf{r},t\right) =\frac{1}{2}\mathbf{\Psi }%
_{\sigma }^{(\alpha )\ast }\cdot \mathbf{\Psi }_{\sigma }^{(-\alpha )}+c.c.,
\label{FieldLikeDensity}
\end{equation}%
where $n_{\sigma }^{(1/2)}$ is essentially equal to $n_{\sigma }^{(0)}$%
except for very broad band signals. The $1$-photon density can be
generalized to describe the photon density in an arbitrary pure state using
the expectation value of the number operator, (\ref{Correlation}).

Systematic investigation of photon position operators and their eigenvectors
clarifies the role of the photon wave function in classical and quantum
optics. The LP wave function defines a positive definite photon number
density and results in photon wave mechanics equivalent to Inagaki's single
photon wave mechanics \cite{Inagaki}. It is related to field based wave
functions through a similarity transformation that preserves eigenvalues and
scalar products. In free space the field (\ref{RSField}) is proportional to
the RS wave function investigated in \cite{BB,BB2,Sipe,Raymer}. The field $%
\mathbf{D}^{(+)}\left( \mathbf{r},t\right) $ is proportional to the Glauber
wave function \cite{Glauber,ScullyBook,LapaireSipe} which gives the
photodetection amplitude for a detector that responds to the electric field 
\cite{Raymer}. While only fields and potentials are locally related to
charge and current sources, Fourier transformation of $\mathbf{k}$-space
probability amplitudes naturally leads to the LP form \cite%
{Eberly,FederovEberly}. The similarity transformation between the
field-potential and LP wave functions makes the choice a matter of
convenience for most purposes.

By the general rules of quantum mechanics the LP wave function is the
probability amplitude to detect a photon at a point in space. It and the
closely related field-potential wave function pair obtained by solution of
MEs are ideally suited to the interpretation of photon counting experiments
using a detector that is small in comparison with the spatial variations of
photon density. It is not subject to limitations based on nonlocalizability,
and coarse graining or restriction to length scales smaller than a wave
length is not required. Exact localization in vacuum requires infinite
energy and is not physically possible, but position eigenvectors provide a
useful mathematical description of photon density . Photon number density is
equivalent to integration over undetected photons in a multiphoton beam. In
an experiment where absorbing particles are placed in a beam, the particles
act as photodetectors which can sense the spin and orbital angular momentum
of the photons. Our formalism justifies the use of positive frequency
Laguerre-Gaussian fields as photon wave functions and gives a rigorous
theoretical basis for extrapolation of their range of applicability from the
many photon to the $1$-photon regime. \ 

\begin{acknowledgement}
The author acknowledges the financial support of the Natural Science and
Engineering Research Council of Canada.
\end{acknowledgement}

\newpage

\end{document}